\documentclass[11pt,a4paper]{article}
\pdfoutput=1

\usepackage{jcappub}
\makeatletter\gdef\@fpheader{}\makeatother
\usepackage[T1]{fontenc}
\usepackage{amsthm}
\usepackage{xcolor}
\usepackage{bm}
\usepackage{enumitem}
\usepackage{mathtools}
\usepackage{tikz}
\usepackage{booktabs}
\usetikzlibrary{calc,arrows.meta,decorations.pathmorphing,positioning}

\newcommand{\thc}{\bm{\theta}^{\mathrm c}}      
\newcommand{\thv}{\bm{\theta}^{\mathrm v}}      
\newcommand{\thb}{\bm{\theta}^{\mathrm b}}      
\newcommand{\PPb}{\overline{\mathrm{PP}}}   

\newcommand{\Cs}{\mathcal{C}}                    
\newcommand{\Vs}{\mathcal{V}}                    
\newcommand{\Pj}[1]{\mathrm{P}_{#1}}             
\newcommand{\Pjperp}[1]{\Pj{#1}^{\!\perp}}       

\newcommand{\F}[1]{\Gamma_{\mathrm{#1}}}         
\newcommand{\Fc}[2]{\Gamma_{\mathrm{#1}\mid\mathrm{#2}}} 

\newcommand{\nip}[2]{\bigl(\,#1\,\big|\,#2\,\bigr)}

\newcommand{\hc}{h_{\mathrm c}}                  

\newcommand{\II}{\mathbb{I}}                     
\DeclareMathOperator{\Tr}{Tr}

\newif\ifvtwohighlight
\vtwohighlighttrue

\newif\ifrevhighlight
\revhighlighttrue

\title{Deglobalising the fit: how to search for physics beyond vacuum General Relativity in LISA catalogues}

\author[1,2]{Francisco Duque}
\author[2]{and Jonathan Gair}

\affiliation[1]{Gravitation Astroparticle Physics Amsterdam (GRAPPA), University of Amsterdam, 1098 XH Amsterdam, The Netherlands}

\affiliation[2]{Max Planck Institute for Gravitational Physics (Albert Einstein Institute),
Am M\"uhlenberg 1, D-14476 Potsdam, Germany}

\emailAdd{francisco.duque@aei.mpg.de}
\emailAdd{jonathan.gair@aei.mpg.de}

\abstract{%
The Laser Interferometer Space Antenna will observe up to $\sim 10^{7}$
overlapping signals, which have to be characterised simultaneously with a
\emph{global} fit. This is a challenge for model comparison since rerunning the
global fit for every model of interest, including deviations from
vacuum General Relativity, is computationally demanding. An alternative is a
two-step post-processing strategy: a first global fit with vacuum models only,
after which selected target sources are reanalysed under a different
model with all other sources subtracted. We show that in this strategy a
beyond-vacuum effect in a target source can be partly absorbed by the
overlapping \textit{nuisance} sources in the first fit, subtracted with them,
and never reach the reanalysis. In the linear-signal approximation, the loss in reported detection
significance is captured by a single number, the \emph{masking ratio}
$\mathcal{R}\in[0,1]$: a deviation the global fit would detect at
$n_\sigma$ is reported by post-processing at $\mathcal{R}\,n_\sigma$, and, for a
single beyond-vacuum parameter, bounds
in null tests are reported too tight by the same factor. We show how to compute
$\mathcal{R}$ from the posterior samples of the vacuum global fit, and give an
algorithm that ranks the nuisance sources by how much of the effect each can
absorb, so that only the most correlated ones need to be refitted with the
target. We validate our method with full Markov-chain Monte
Carlo inference on a toy model of 21 overlapping chirping sinusoids, one of
which carries a small non-linear chirp that represents the missed physics.
Using our post-processing recipe, we are able to recover the global-fit result on the beyond-vacuum parameter to
within $0.15\sigma$ by refitting only 4 of the 20 nuisance sources. This method
can serve as a basis for tests of General Relativity in realistic LISA mock
datasets.}

\begin{document}
\maketitle

\section{Introduction}

The current network of ground-based gravitational-wave (GW) detectors finds a new event every couple of days of operation~\cite{LIGOScientific:2026sit, LIGOScientific:2026wfs}.
As instrument sensitivity improves,
signals will begin to overlap and
will have to be fitted jointly when sufficiently correlated.
This is not a problem for the LIGO-Virgo-KAGRA network,
but it is expected to happen for next-generation ground-based detectors, such as the Einstein Telescope and Cosmic Explorer~\cite{ET:2025xjr, Reitze:2019iox, Iacovelli:2022bbs, Janquart:2022fzz}, where
a typical binary neutron star signal will have tens of other signals overlapping it,
and two events will merge within seconds of each other tens of thousands of times per year~\cite{Samajdar:2021egv}.
The challenge will be even greater for the space-based Laser Interferometer Space Antenna (LISA), scheduled to launch in 2035~\cite{LISA:2024hlh}.
LISA will probe the millihertz band, where many sources are long-lived:
$\sim 10^{7}$ Galactic binaries, of which $\sim 10^{4}$ are individually resolvable while the rest contribute to an unresolved foreground~\cite{LISA:2024hlh};
up to hundreds of massive BH binaries with total masses of $10^{4}$--$10^{7}\, M_\odot$~\cite{Klein:2015hvg};
up to thousands of extreme-mass-ratio inspirals~\cite{Babak:2017tow, Speri:2026ade}, in which a stellar-mass compact object spirals into a massive BH while completing up to $\sim 10^{5}$ orbital cycles over months to years;
a few stellar-mass BH binaries in their early inspiral, years before they enter the band where ground-based detectors may observe them~\cite{Klein:2022rbf};
and possibly stochastic GW backgrounds of astrophysical origin (e.g. unresolved extragalactic stellar-mass binaries)~\cite{Babak:2023lro, Pozzoli:2024wfe, Pozzoli:2023kxy} or cosmological origin (e.g. phase transitions in the early Universe)~\cite{LISACosmologyWorkingGroup:2022jok}.
Adding more complexity to the problem, LISA's noise is not known a priori and is non-stationary, with gaps and glitches~\cite{Boumerdassi:2025gvf, Burke:2025fvl, Muratore:2025knh, Burke:2025bun}, so it must also be inferred jointly with the signals.

This sheer number of overlapping signals requires a \textit{global fit} to
extract all sources from the data simultaneously~\cite{Katz:2024oqg,
Littenberg:2023xpl, PhysRevD.111.103014, Strub:2024kbe, Alvey:2023naa, Agrawal:2026tbd}. Global fit algorithms are under active development using increasingly more realistic LISA mock
datasets~\cite{MockLISADataChallengeTaskForce:2009wir}. The current plan to
tackle real data is for the first analysis to use waveform models assuming only
two-body, vacuum General Relativity (GR) for all binary systems. Public data products will then be released that enable the scientific community to pursue science investigations that go beyond those initial catalogues
~\cite{LISA:SIRD}, for example using more accurate vacuum waveform  models,
models with beyond-GR effects~\cite{Barausse:2020rsu, LISA:2022kgy, LISAConsortiumWaveformWorkingGroup:2023arg}, or astrophysical
environmental effects~\cite{Yunes:2011ws, Barausse:2014tra}. How this will be achieved in practice is an open question. Rerunning a full
global fit for every model the community might want to test is infeasible, both
in complexity and computational cost. Instead, it is expected that the
two-body, vacuum models will be accurate enough that any deviation from them is
small, so that the public data products can be post-processed, reanalysing only a
few \emph{target} sources of interest~\cite{Kejriwal:2025upp}. To test GR, for
example, the logical choice would be to pick sources with large signal-to-noise ratio
(SNR), and/or with an anomalous accumulation of SNR~\cite{Kejriwal:2025upp, Copparoni:2025vty}.

However, correlations between sources can mask physical effects already in the
initial vacuum global fit and remain hidden in post-processing. Run
under the vacuum hypothesis, that first fit can absorb any beyond-vacuum effect
into its parameters. Part of the effect goes into the target source's own
``normal'', vacuum parameters, and a post-processing refit can (in principle) undo
this. Another part is absorbed by overlapping \emph{nuisance} sources, which are
then subtracted from the data having already ``eaten'' part of the effect, which
never reaches the post-processing step, and a reanalysis of the target source
alone cannot recover it (note the same correlations matter when no deviation is
present). Post-processing then quotes an uncertainty on the beyond-vacuum
parameters that treats the nuisance sources as perfectly known, when in fact they
were fitted and could themselves have absorbed a deviation had one been there. The
upper bound it reports is therefore tighter than what it would be in the global fit. The propagation of biases between overlapping sources has been studied
in~\cite{Antonelli:2021vwg}, within the same linear-signal treatment we use here.
That work studied how unfitted signals, inaccurate waveforms and imperfectly
subtracted sources bias the recovered parameters, including a
treatment of the unresolved foreground, and proposed iteratively correcting the
peaks of a source-by-source analysis towards the global-fit answer. Here, we are instead
concerned with how much
of the evidence for a beyond-vacuum effect the two-step analysis reports, and
which sources have to be refitted jointly for it to be recovered.

To achieve this, it would be useful to have a computationally cheap  method to identify,
after the first vacuum global fit, which sources are most able to mask a given
unmodelled effect in another source, so that these can be analysed simultaneously
in a `down-scaled' reanalysis, a joint refit of the target and those sources only. This is what we construct
here. As we will show, under the linear-signal approximation, the difference in the
detection significance that the two strategies (global fit \textit{vs}
post-processing) report for a deviation from vacuum GR can be captured by the
\emph{masking ratio} $\mathcal{R}$, which is a single number between 0 and 1. When a deviation is present, it is the
factor multiplying the detection significance, so that a deviation the global fit
would establish at $n_\sigma$ is reported by post-processing at
$\mathcal{R}\,n_\sigma$. It is also the factor by which the upper
bound is overconfidently quoted when no effect is in the data.  We
also show a practical method to compute this ratio directly from the vacuum
global-fit posterior samples, and an algorithm that, given a target source and a
beyond-vacuum model, ranks the nuisance sources by how much of the effect each is
able to absorb. Our treatment can also be applied to evaluate
waveform systematics, tracking how
they propagate to errors in the other overlapping sources and how they can lead to
a false-positive deviation from GR. We validate our method with full
Markov-chain Monte Carlo inference on a toy model of 21 chirping sinusoids.

This problem is also related to the \emph{fundamental bias} of tests of GR with
single sources~\cite{Yunes:2009ke, Vallisneri:2012qq, Vallisneri:2013rc}, where the
parameters of a source absorb part of any unmodelled effect, which biases them,
and only what they cannot absorb can be detected. If the bias exceeds the
statistical errors before the effect becomes detectable, it is
\emph{stealthy}~\cite{Cornish:2011ys,Vallisneri:2013rc}. We extend this picture
to overlapping sources, whose parameters can absorb the effect as well, and show
that post-processing adds an additional bias of its own. After the biased neighbours are
subtracted, the target-only reanalysis reports a weaker effect, with
uncertainties that are too narrow, even when the global fit would have detected
it.

The paper is organised as follows: in Sec.~\ref{sec:recipe}, we summarise our
results as a practical recipe. In Sec.~\ref{sec:setup}, we set up the notation and give a geometric and statistical picture of the
problem. Secs.~\ref{sec:StrategyGF} and~\ref{sec:StrategyPP} derive the detection statistics of the global fit and of post-processing, including a variant that accounts for the uncertainty of the subtracted sources. In Sec.~\ref{sec:comparison}, we compare them, and define the masking and
null ratios that quantify the loss in the detection significance reported by the
post-processing strategy. In
Sec.~\ref{sec:practical} we show how to compute these ratios from the vacuum global-fit posterior and how to select the sources to refit in order to minimise the loss in detection significance in post-processing. We validate the method with full
Markov-chain Monte Carlo inference on a toy model in Sec.~\ref{sec:ToyModel}, and discuss its applicability and limitations in Sec.~\ref{sec:Disc}.

\section{Recipe for tests of vacuum General Relativity in post-processing}
\label{sec:recipe}

For readers mainly interested in applying our method, we first summarise our
results in the form of a recipe for post-processing a target source after a
vacuum global fit. Each step refers to
the section where it is derived, and every quantity it needs can be computed
from the posterior samples of the vacuum global fit without rerunning it.
\begin{enumerate}
\item From the posterior samples of the vacuum global fit, compute the masking ratio $\mathcal{R}$ and the
  null ratio $\mathcal{R}_0$ (Sec.~\ref{sec:comparison}). The smaller of
  $\mathcal{R}$ and $\sqrt{\mathcal{R}_0}$ bounds the loss for an effect of any
  size (Eq.~\eqref{eq:Rmin}). For several beyond-vacuum parameters and no specific model, use instead the
worst case over all directions, $\sqrt{w_{\min}}$ of Eq.~\eqref{eq:Rbracket},
where $w_{\min}$ is the smallest of the weights of Eq.~\eqref{eq:chi2pp}.
\item If the number computed in Step 1 is close enough to 1, subtract all nuisance sources and
  refit the target source alone under the new model being studied. How close is close enough depends on how much detection
  significance one is willing to give up. For example, a value of $0.9$ obtained in Step 1 means
  that a deviation the global fit would detect at $5\sigma$ is still reported at $4.5\sigma$ or more, and that bounds in null tests are at most $10\%$ too
  tight.
\item If the number computed in Step 1 is below the determined threshold, rank the nuisance sources with the algorithm of
  Sec.~\ref{sec:selection}, and add them to a set $S$ until the same number,
  computed for the residual masking (Eqs.~\eqref{eq:factorise}
  and~\eqref{eq:R0S}) is close enough to one. If the effect is large,
  sweep the expansion point over the allowed range of the beyond-vacuum
  parameters, refitting the vacuum parameters at each value, and take the
  union of the selected sources, or re-evaluate the
  ranking at the best fit of each refit until the selected set no
  longer changes.
\item Subtract only the sources outside $S$, and refit the target sources jointly with
  the sources in $S$.
\end{enumerate}
The nuisance sources can be subtracted in two ways. Subtracting them at their
best fit is cheaper and gives detection significances
closer to those of the global fit, so it is the better choice for detecting a
deviation. Averaging over the posterior of the first fit instead
costs one refit per posterior draw and up to a
factor $\sqrt{2}$ in detection significance, but brings the upper limits of a
null test closer to those of the global fit, so it is the better choice for
setting bounds.

\section{Setup and Notation}
\label{sec:setup}

We take the observed GW data stream $d$ to be
\begin{equation}
d=n+h(\thv_0,\thb_0)+\hc(\thc_0)+\delta m ,
\label{eq:data}
\end{equation}
where $n$ is zero-mean Gaussian noise with covariance
$C_n=\langle n \, n^{\dagger}  \rangle$ and $\langle\,\cdot\,\rangle$ denotes an
average over noise realisations. Here, $h(\thv,\thb)$ and $\hc(\thc)$ are the
waveform models used in the analysis, so that the full template is
\begin{equation}
m(\thc,\thv,\thb)=h(\thv,\thb)+\hc(\thc) ,
\label{eq:template}
\end{equation}
and we separate the modelled signals into two classes:
\begin{itemize}[leftmargin=1.5em]
\item $h$: the target source(s) of interest, possibly carrying a beyond-vacuum effect;
\item $\hc$: nuisance (or confusion) sources, without a beyond-vacuum effect.
\end{itemize}
The parameter space is also divided into three blocks:
\begin{itemize}[leftmargin=1.5em]
\item $\thb$: the beyond-vacuum parameters of the source(s) of interest,
with $\thb=\bm 0$ recovering vacuum;
\item $\thv$: the vacuum parameters of the target source(s);
\item $\thc$: the nuisance-source parameters.
\end{itemize}
A subscript ``$0$'' denotes the true value of a parameter and ``$\mathrm{bf}$''
a best-fit value under a given hypothesis. The other term in Eq.~\eqref{eq:data}, $\delta m$, is the model error, representing all signal components of the data that the template
does not include when evaluated at the true parameters. These  errors could come from unmodelled (quiet) sources, waveform systematics, or instrumental artefacts the noise model $C_n$ does not describe, like glitches.

We assume a Gaussian likelihood so that
$-2\log L=\nip{d-m}{d-m}+\mathrm{const}$, where $\nip{\,\cdot\,}{\,\cdot\,}$
is the noise-weighted inner product
\begin{equation}
\nip{a}{b}= a^{\dagger}\,C_n^{-1}\,b .
\label{eq:inner}
\end{equation}

We caution the reader that the next subsections are particularly heavy on notation, but introducing it makes the presentation of our results later on much more tractable.

\subsection{Geometric intuition}
\label{sec:geometry}

Before delving into any statistical analysis, we can use the inner product~\eqref{eq:inner} to build a geometric picture of the problem.  Varying the template parameters traces out a manifold in data space. The vacuum
(sub)manifold is defined by $\thb=\bm 0$, so it contains everything a vacuum
global fit can represent. Near a reference
point in parameter space, the model is characterised by
its tangent directions $\partial_{\rm X}m$, so its tangent space is
\begin{equation}
\Vs\oplus\Cs=\mathrm{span}\{\partial_{\thv}m,\ \partial_{\thc}m\} ,
\label{eq:tangent}
\end{equation}
where $\Vs$ is spanned by the vacuum directions of the target source and $\Cs$ by
the nuisance-source directions. We can project any vector onto this space with
\begin{equation}
\Pj{\Vs\oplus\Cs}=\!\!\sum_{{\rm X},{\rm Y}\in\{{\rm c},{\rm v}\}}\!\!
|\partial_{\rm X}m)\;\bigl[\Gamma_{\rm cv,cv}^{-1}\bigr]_{\rm XY}\;(\partial_{\rm Y}m| ,
\qquad
\Pjperp{\Vs\oplus\Cs}=\II-\Pj{\Vs\oplus\Cs} ,
\label{eq:projdef}
\end{equation}
where $\Gamma_{\rm cv,cv}$ is the $\{\rm c,v\}$ sub-block of
\begin{equation}
\Gamma_{\rm XY} = \nip{\partial_{\rm X} m}{\partial_{\rm Y} m} \, .
\label{eq:FM}
\end{equation}

A beyond-vacuum effect displaces the model in data space by
$(\partial_{\thb}m)\,\thb$, which we assume is small enough that the curvature of the vacuum
submanifold is negligible. The part of this displacement along
$\Vs\oplus\Cs$ can be reproduced by adjusting the vacuum and nuisance-source
parameters, and is therefore indistinguishable from vacuum. Only the orthogonal
component is in principle observable, which has length
\begin{equation}
\bigl\|\,\Pjperp{\Vs\oplus\Cs}(\partial_{\thb}m)\,\thb\,\bigr\|^{2}
=(\thb)^{T}\bigl[\F{bb}-\F{b,cv}\,\Gamma_{\rm cv,cv}^{-1}\,\F{cv,b}\bigr]\thb
=(\thb)^{T}\Fc{b}{cv}\,\thb ,
\label{eq:geomGamma}
\end{equation}
with
\begin{align}
 \F{b,cv}&=\begin{pmatrix}\F{bc}&\F{bv}\end{pmatrix} , \nonumber \\
 \F{cv,b}&=\begin{pmatrix}\F{cb}\\ \F{vb}\end{pmatrix} = (\F{b,cv})^{T} .\
\end{align}
The bracket defines the \emph{conditional} matrix
$\Fc{b}{cv}$, whose statistical meaning is given in Sec.~\ref{sec:statistics}. We make this more precise below, but intuitively, to
claim a deviation from vacuum GR we need this length to be large, so
\begin{equation}
\bigl\|\,\Pjperp{\Vs\oplus\Cs}(\partial_{\thb}m)\,\thb\,\bigr\|^{2}
= (\thb)^{T}\Fc{b}{cv}\,\thb \gtrsim \mathcal{O}(1)  \, ,
\label{eq:GeometricDetCriteria}
\end{equation}
where the number on the right-hand side is set by the confidence level at which the
deviation is claimed ($2\sigma$, $3\sigma$, etc). In practice, the criterion is applied to the recovered best fit,
$\thb_{\rm bf}=\thb_0+\Delta\thb$, which noise and model error displace from the
true value. This displacement depends on which parameters are fitted jointly
with $\thb$, so it differs between the global fit and post-processing, and we will
compute it for each strategy in Secs.~\ref{sec:StrategyGF}
and~\ref{sec:StrategyPP}, respectively.

The same construction applies to any subspace. In the simplest post-processing strategy, the nuisance sources are subtracted from the data as if they were perfectly known and only the target source is reanalysed. Therefore, only $\Vs$ is then available to absorb the
effect in the reanalysis of the residual, and projecting out $\Vs$ alone gives a second conditional matrix $\Fc{b}{v}$
\begin{equation}
\Fc{b}{v}=\nip{\partial_{\thb}m}{\Pjperp{\Vs}\,\partial_{\thb}m} ,
\qquad
\Fc{b}{cv}=\nip{\partial_{\thb}m}{\Pjperp{\Vs\oplus\Cs}\,\partial_{\thb}m} \, .
\label{eq:geomboth}
\end{equation}
Since $\Vs\subseteq\Vs\oplus\Cs$, intuitively projecting out the larger subspace can only
shorten the vector. Their difference is what
nuisance sources are able to absorb.
This has two consequences for
post-processing (which we make precise in Sec.~\ref{sec:StrategyPP}). First,
because the vector is longer, the posterior on $\thb$ is narrower than in the
global fit, so when no effect is present post-processing is overconfident in the
upper limit it sets. Second, when an effect is present, the part of it lying
along $\Cs$ is absorbed by the nuisance sources in the first fit and subtracted
with them, so it never reaches the reanalysis, and the recovered value is biased
towards vacuum. The extra length that makes post-processing look more sensitive
is therefore precisely the part of the effect that the nuisance sources have
already absorbed.

\begin{figure}[t]
\centering
\includegraphics[width=\textwidth]{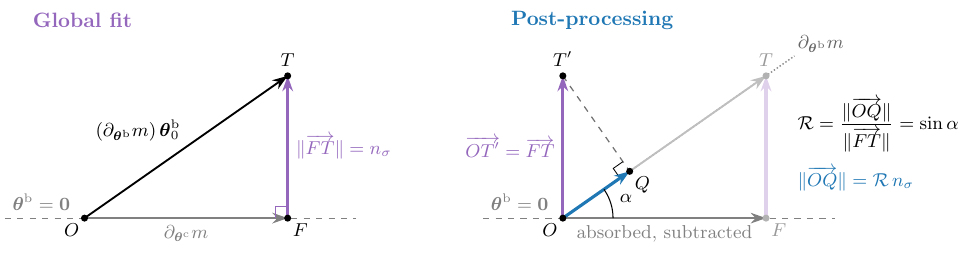}
\caption{Geometric picture of the masking effect in post-processing for a single beyond-vacuum parameter (without noise). Lengths correspond to signal-to-noise ratios measured with the noise-weighted inner product~\eqref{eq:inner}. The dashed line contains the vacuum templates
($\thb=\bm 0$), along which the nuisance-source parameters $\thc$ move the
signal. The vacuum parameters of the target, which both strategies refit, are
projected out. The data point $T$ differ from the vacuum point $O$ by the beyond-vacuum
effect $(\partial_{\thb}m)\,\thb_0$, and $F$ is the vacuum template closest to
$T$. \emph{Left:} $\|\overrightarrow{FT}\|^{2}$ is twice the log-likelihood ratio
between the best beyond-vacuum and vacuum fits, so the global fit detects the
deviation at $n_\sigma=\|\overrightarrow{FT}\|$. \emph{Right:} post-processing
subtracts the nuisance sources together with the 
$\overrightarrow{OF}$ signal portion they absorbed in the vacuum fit, and refits the target source on
the residual $T'$ along the beyond-vacuum direction, recovering $Q$. The
deviation from vacuum is then quoted at $\|\overrightarrow{OQ}\|=\mathcal{R}\,n_\sigma$,
with $\mathcal{R}=\sin\alpha$~\eqref{eq:R_geometric}.}
\label{fig:geometry}
\end{figure}

For a single beyond-vacuum parameter, all these ideas follow from trigonometry in the plane of Fig.~\ref{fig:geometry}. With the inner
product of Eq.~\eqref{eq:inner}, signals are vectors whose lengths are
signal-to-noise ratios, and $-2\log L$ is, up to a constant, the squared distance
between the data and the template, so at linear order a maximum-likelihood fit is
the orthogonal projection of the data onto the directions the template can span.
Both strategies refit the vacuum parameters of the target source, so we work
orthogonally to $\Vs$, with the beyond-vacuum direction
$\tilde u=\Pjperp{\Vs}\,\partial_{\thb}m$, of squared length
$\nip{\tilde u}{\tilde u}=\Fc{b}{v}$, and the nuisance-source directions
projected in the same way, $\Pjperp{\Vs}\Cs$. Without noise, the data sit at $T$,
displaced from the vacuum point $O$ by $\overrightarrow{OT}=\tilde u\,\thb_0$, and we call $\alpha$ the angle between $\tilde u$ and
$\Pjperp{\Vs}\Cs$. The vacuum best fit recovers the closest point, $F$, while the
beyond-vacuum fit recovers $T$ exactly, so the two values of $-2\log L$ differ
by $\|\overrightarrow{FT}\|^{2}=\|\overrightarrow{OT}\|^{2}\sin^{2}\alpha$. This
is twice the logarithm of the maximum-likelihood ratio between the two
hypotheses, so its square
root is the detection significance, i.e. how many ``sigmas'' we are away from vacuum $n_\sigma=\|\overrightarrow{FT}\|$ (which also
sets the Bayes factor
used in Bayesian inference for model comparison). Comparing with Eq.~\eqref{eq:geomGamma} also
gives $\sin^{2}\alpha=\Fc{b}{cv}/\Fc{b}{v}$.

In post-processing, the vacuum fit absorbs $\overrightarrow{OF}$ into the
nuisance sources, which are subtracted, so the reanalysis sees only the residual
$\overrightarrow{OT'}=\overrightarrow{FT}$. With the nuisance sources fixed, the
reanalysis can only move along $\tilde u$, and it lands on the projection $Q$ of
$T'$, and
$\|\overrightarrow{OQ}\|=\|\overrightarrow{FT}\|\sin\alpha$. Then, by Pythagoras theorem
the two post-processing fits differ in $-2\log L$ by
$\|\overrightarrow{OT'}\|^{2}-\|\overrightarrow{T'Q}\|^{2}=\|\overrightarrow{OQ}\|^{2}$.
Post-processing therefore reports the deviation at
$\|\overrightarrow{OQ}\|=n_\sigma\sin\alpha$. It recovers the value
$\|\overrightarrow{OQ}\|/\|\tilde u\|=\thb_0\sin^{2}\alpha$, shrunk towards
vacuum and gives the uncertainty $1/\|\tilde u\|=\Fc{b}{v}^{-1/2}$, narrower by
$\sin\alpha$ than that of the global fit, which can only use the component of
$\tilde u$ perpendicular to the nuisance directions,
$1/(\|\tilde u\|\sin\alpha)=\Fc{b}{cv}^{-1/2}$. All these three quantities are set by a single
number, the \emph{masking ratio}
\begin{equation}
\mathcal{R}\equiv\frac{\|\overrightarrow{OQ}\|}{\|\overrightarrow{FT}\|}
=\sin\alpha=\left(\frac{\Fc{b}{cv}}{\Fc{b}{v}}\right)^{1/2}\in[0,1] \, .
\label{eq:R_geometric}
\end{equation}
A deviation that the global fit detects at $n_\sigma$ is reported by
post-processing at $\mathcal{R}\,n_\sigma$, its value is recovered shrunk by
$\mathcal{R}^{2}$, and its error bar is too narrow by $\mathcal{R}$. The masking ratio is 1 when the
beyond-vacuum direction is perpendicular to everything the nuisance sources can
absorb, and tends to zero when they can mimic it completely.

\subsection{Statistical interpretation}
\label{sec:statistics}

We now give a more statistical meaning to each object in the geometric picture above.
The matrix $\Gamma$~\eqref{eq:FM} is the \textit{Fisher information} matrix~\cite{Cutler:2007mi, Vallisneri:2007ev}. It is the curvature of the (Gaussian) log-likelihood around the truth in the linear-signal approximation, in which the template is expanded to first order in the parameter displacement so that the log-likelihood becomes quadratic. In the same approximation, the posterior is Gaussian with covariance $\Sigma=\Gamma^{-1}$, so that correlations between parameters and the measurement uncertainty in the quantity $X$ are
\begin{equation}
\sigma_{\rm X}^{2}=(\Gamma^{-1})_{\rm X} , \qquad
\mathrm{corr}(\theta_{\rm X},\theta_{\rm Y})
 =\frac{(\Gamma^{-1})_{\rm XY}}
        {\sqrt{(\Gamma^{-1})_{\rm X}\,(\Gamma^{-1})_{\rm Y}}} .
\label{eq:cov}
\end{equation}
Noise and model error do not affect the covariance at leading order, but bias the recovered
parameters by the Cutler--Vallisneri bias~\cite{Cutler:2007mi}
\begin{equation}
\Delta {\theta^{\rm X}}
   =(\Gamma^{-1})_{\rm XY}\,\nip{\partial_{\rm Y} m}{\delta m+n} \, ,
\label{eq:cvbias}
\end{equation}
which is  the projection of the residual (model error and noise) onto the tangent vector of the
corresponding parameter direction. The noise contribution averages to zero over
many realisations, but has non-zero variance. The noise-induced fluctuations are statistically consistent with the width of the posterior distributions.

Projecting out a block of directions $\rm Z$, as done above in the geometric picture, is what in statistics
is called \emph{marginalisation}. It corresponds to integrating out $\rm Z$ while retaining the
uncertainty it induces on the remaining blocks. It gives the conditional Fisher
matrix
\begin{equation}
\Gamma_{\mathrm{XY}\mid\mathrm{Z}}
= \F{XY}-\F{XZ}\,\F{ZZ}^{-1}\,\F{ZY} ,
\label{eq:conddef}
\end{equation}
with $\Fc{X}{Z}$ shorthand for the $\rm X=Y$ case. Marginalising over two blocks at once factorises,
\begin{equation}
\Gamma_{\mathrm{X}\mid\mathrm{YZ}}
= \Fc{X}{Z}-\Gamma^T_{\mathrm{YX}\mid\mathrm{Z}}\,
  \Gamma_{\mathrm{YY}\mid\mathrm{Z}}^{-1}\,
  \Gamma_{\mathrm{YX}\mid\mathrm{Z}} ,
\label{eq:chainrule}
\end{equation}
so that marginalising over $\{\mathrm Y,\mathrm Z\}$ is the same as marginalising
over $\rm Z$ and then $\rm Y$ (or vice versa). Marginalisation can only lead to a loss of information. Note also that since
$\Gamma$ is positive-definite, $x^T \Gamma_{\rm X\mid Z}\, x \leq x^T
\Gamma_{\rm X}\, x$ for every $x$, which we write
$\Gamma_{\rm X}\succeq \Gamma_{\rm X\mid Z}$.

The inverse of a two-block Fisher matrix is (in the same conditional notation)
\begin{equation}
\begin{pmatrix}\F{Z}&\F{ZX}\\\F{XZ}&\F{X}\end{pmatrix}^{-1}
= \begin{pmatrix}
\Fc{Z}{X}^{-1} & -\Fc{Z}{X}^{-1}\,\F{ZX}\,\F{X}^{-1}\\[4pt]
-\Fc{X}{Z}^{-1}\,\F{XZ}\,\F{Z}^{-1} & \Fc{X}{Z}^{-1}
\end{pmatrix}.
\label{eq:InvBlock}
\end{equation}
Here, $\rm X$ and $\rm Z$ can each stand for a group of parameters, so the same rule
inverts a matrix with more than two blocks once its rows and columns are split into
the respective subsets. Taking $\rm Z=\{c,v\}$ and $\rm X=b$ shows
$\Fc{b}{cv}=[(\Gamma^{-1})_{\rm bb}]^{-1}$. For a single beyond-vacuum GR
parameter $\Fc{b}{cv}=1/\sigma_{\thb}^{2}$ with $\sigma_{\thb}$ its fully
marginalised uncertainty, and $(\thb)^{T}\Fc{b}{cv}\,\thb=(\thb/\sigma_{\thb})^{2}$
is simply the distance from vacuum in units of that uncertainty. The geometric
criterion~\eqref{eq:GeometricDetCriteria} is then the statement that a confident
deviation from GR requires a shift (much) larger than the statistical uncertainty in the measurement of the beyond-vacuum GR parameter.

\section{Global Fit strategy detection statistic}
\label{sec:StrategyGF}

The detectability criterion of Eq.~\eqref{eq:GeometricDetCriteria} is what one
obtains in the full global fit, in which the beyond-vacuum GR parameters are included
in the model from the start. We rederive it here with the statistical machinery of
Sec.~\ref{sec:statistics}, which also gives us the bias $\Delta\thb_{\mathrm{GF}}$ and the effect of systematic errors in the model.

In the global fit strategy, the model used to analyse the data is
\begin{equation}
m^{\mathrm{GF}}(\thc,\thv,\thb)=\hc(\thc)+h(\thv,\thb).
\end{equation}
The Fisher matrix splits into three blocks
\begin{equation}
\Gamma^{{\rm GF}}=
\begin{pmatrix}\F{cc}&\F{cv}&\F{cb}\\\F{vc}&\F{vv}&\F{vb}\\\F{bc}&\F{bv}&\F{bb}\end{pmatrix}.
\label{eq:GFFisher}
\end{equation}
The off-diagonal terms, $\Gamma_{\rm cb}$ and $\Gamma_{\rm vb}$,
quantify how much the nuisance sources and the vacuum
parameters can absorb the beyond-vacuum effect.
Marginalising over both, using Eq.~\eqref{eq:chainrule}, we have
\begin{equation}
\Fc{b}{cv}=\Fc{b}{v}-\Gamma_{\mathrm{cb}\mid\mathrm{v}}^{T}\Fc{c}{v}^{-1}\Gamma_{\mathrm{cb}\mid\mathrm{v}} \, .
\label{eq:cGamma}
\end{equation}
This joint Fisher matrix over a target source and its overlapping neighbours,
and the conditioning of its target block,
were used in Ref.~\cite{Antonelli:2021vwg} to quantify how fitting the neighbours jointly
degrades the precision on the target’s parameters.

The best-fit value $\thb_{\rm bf}=\thb_0+\Delta\thb_{\mathrm{GF}}$ is at a
squared distance from vacuum, in units of its uncertainty, of
$2S^{\mathrm{GF}}$, with
\begin{equation}
S^{\mathrm{GF}}=\tfrac12\bigl(\thb_0+\Delta\thb_{\mathrm{GF}}\bigr)^{T}
\Fc{b}{cv}\bigl(\thb_0+\Delta\thb_{\mathrm{GF}}\bigr) \, .
\label{eq:Sgf_def}
\end{equation}
To compute $\Delta\thb_{\mathrm{GF}}$, we apply Eq.~\eqref{eq:cvbias} with
residual $n+\delta m$, inverting Eq.~\eqref{eq:GFFisher} using
Eq.~\eqref{eq:InvBlock} with the blocks grouped as $\rm Z=\{c,v\}$ and
$\rm X=b$, so that the beyond-vacuum GR block of the inverse is $\Fc{b}{cv}^{-1}$. We find
\begin{equation}
\Delta\thb_{\mathrm{GF}}=\Fc{b}{cv}^{-1}(\Pjperp{\Vs\oplus\Cs}(\partial_{\thb}m)|n + \delta m) =\Fc{b}{cv}^{-1}\bm Z^{\mathrm{GF}},
\qquad
\bm Z^{\mathrm{GF}}=\bm Z_n^{\mathrm{GF}}+\bm Z_{\delta m}^{\mathrm{GF}},
\label{eq:bGF}
\end{equation}
with
\begin{equation}
\bm Z_{\bullet}^{\mathrm{GF}}=
\nip{\partial_{\thb}m}{\bullet}
-\begin{pmatrix}\F{bc}&\F{bv}\end{pmatrix}
 \begin{pmatrix}\F{cc}&\F{cv}\\\F{vc}&\F{vv}\end{pmatrix}^{-1}
 \begin{pmatrix}\nip{\partial_{\thc}m}{\bullet}\\ \nip{\partial_{\thv}m}{\bullet}\end{pmatrix} \, ,
\label{eq:Zsplit_gf}
\end{equation}
and $\bullet=(n, \delta m)$. Geometrically, $\bm Z_{\bullet}^{\mathrm{GF}}$ is
the overlap of the residual with $\Pjperp{\Vs\oplus\Cs}(\partial_{\thb}m)$. The operator $\Pjperp{\Vs\oplus\Cs}$ projects into the space not spanned by changes to the vacuum and nuisance source parameters, while $\{\partial_{\thb}m\}$ spans the space accessible by changing the beyond-vacuum parameters. This implies that the shift in the beyond-vacuum parameters is driven by the part of the residual that can be absorbed by changes to the beyond-vacuum parameters and {\it not} by changes to the vacuum and nuisance source parameters. The noise
piece $\bm Z_n^{\mathrm{GF}}$ has zero mean and covariance $\Fc{b}{cv}$ by
Eq.~\eqref{eq:InvBlock}. Thus, averaging $S^{\mathrm{GF}}$ over noise, the term linear in
$\bm Z_n^{\mathrm{GF}}$ vanishes and the quadratic one gives
$\tfrac12\Tr[\Fc{b}{cv}^{-1}\Fc{b}{cv}]=\tfrac12 N_{\mathrm b}$, with
$N_{\mathrm b}=\dim\thb$, so that
\begin{equation}
\langle S^{\mathrm{GF}}\rangle
=\tfrac12\bigl[(\thb_0)^{T}\Fc{b}{cv}\thb_0+N_{\mathrm b}\bigr]
+(\thb_0)^{T}\bm Z_{\delta m}^{\mathrm{GF}}
+\tfrac12\,(\bm Z_{\delta m}^{\mathrm{GF}})^{T}\Fc{b}{cv}^{-1}\bm Z_{\delta m}^{\mathrm{GF}}.
\label{eq:maskgf_u}
\end{equation}
The first term is the signal contribution in the fully marginalised metric,
the cross term an interference between model error and signal, and the last one
is present even in vacuum ($\thb_0=\bm 0$), where it can lead to a false-positive
deviation from vacuum GR.
Note that for no model errors,
$\delta m=0$, the detection statistic becomes
\begin{equation}
\langle S^{\mathrm{GF}}\rangle
=\tfrac12\bigl[(\thb_0)^{T}\Fc{b}{cv}\thb_0+N_{\mathrm b}\bigr] \, ,
\label{eq:maskgf}
\end{equation}
while with model errors, the three terms of Eq.~\eqref{eq:maskgf_u} form a perfect square by defining
the effective beyond-vacuum parameter (which is the noise average of the recovered best fit value)
\begin{equation}
\thb_{\mathrm{eff}}= \langle \thb_{\rm bf} \rangle = \thb_0+\Fc{b}{cv}^{-1}\bm Z_{\delta m}^{\mathrm{GF}} \, ,
\label{eq:thetaeff}
\end{equation}
in terms of which noise-averaged detection statistic becomes
\begin{equation}
\langle S^{\mathrm{GF}}\rangle
=\tfrac12\bigl[(\thb_{\mathrm{eff}})^{T}\Fc{b}{cv}\thb_{\mathrm{eff}}
+N_{\mathrm b}\bigr] .
\label{eq:maskgf_eff}
\end{equation}
%

\subsection{Maximum likelihood detection significance}
\label{sec:significanceGF}

$S^{\mathrm{GF}}$ is minus the argument of the Gaussian likelihood evaluated for vacuum-GR and, as such, can be directly interpreted as a significance of a departure from vacuum GR, in terms of (the sum over beyond-vacuum-GR parameters of the squares of) the number of standard deviations the best-fit parameters are away from zero. However, it can also be interpreted as a Bayes Factor or as the standard test statistic for identifying a deviation from vacuum-GR. If we maximise the
(Gaussian) likelihood over all parameters \textit{vs}
with $\thb$ held at $\bm 0$ (the null hypothesis), the two best-fit residuals differ by
\begin{equation}
\nip{r_{\rm vac}}{r_{\rm vac}}-\nip{r_{\rm bf}}{r_{\rm bf}}
=(\thb_{\rm bf})^{T}\Fc{b}{cv}\,\thb_{\rm bf}
=2S^{\mathrm{GF}} \, ,
\label{eq:LLR}
\end{equation}
so $S^{\mathrm{GF}}$ is the log-likelihood ratio between the beyond-vacuum and
vacuum hypotheses.  It follows a non-central $\chi^{2}$ distribution
\begin{equation}
2S^{\mathrm{GF}}\ \sim\ \chi^{2}_{N_{\mathrm b}}\bigl(\lambda^{\mathrm{GF}}\bigr) ,
\qquad
\lambda^{\mathrm{GF}}=(\thb_{\mathrm{eff}})^{T}\Fc{b}{cv}\,\thb_{\mathrm{eff}} ,
\label{eq:chi2}
\end{equation}
with $N_{\mathrm b}$ degrees of freedom and
mean $N_{\mathrm b}+\lambda^{\mathrm{GF}}$. A detection is then claimed at significance $\sqrt{2S^{\mathrm{GF}}}$, which for a
single beyond-vacuum parameter is the recovered value in units of its uncertainty,
$|\thb_{\rm bf}|/\sigma_{\thb}$ with $\sigma_{\thb}=\Fc{b}{cv}^{-1/2}$. For a
strong beyond-vacuum GR effect, it averages to $\sqrt{\lambda^{\mathrm{GF}}}$. Conversely, in vacuum (and without model errors) the distribution is central, so a false-alarm probability
$\alpha=\mathrm{Prob}(\chi^{2}_{N_{\mathrm b}}(0)>2S_{*})$ fixes the threshold
$2S_{*}$, or equivalently $n_{\sigma}=\sqrt{2}\,\mathrm{erfc}^{-1}(\alpha)$ sigmas. For one beyond-vacuum parameter the two coincide,
$2S_{*}=n_{\sigma}^{2}$, so three sigma is $2S_{*}=9$ and five sigma
$2S_{*}=25$. Note that the threshold at fixed $\alpha$ grows with $N_{\mathrm b}$.

Also, since $\lambda^{\mathrm{GF}}$ is built from $\thb_{\mathrm{eff}}$ instead of
$\thb_0$, model error alone makes the distribution non-central in vacuum, with
non-centrality
$\lambda_{\delta m}=(\bm Z_{\delta m}^{\mathrm{GF}})^{T}\Fc{b}{cv}^{-1}\bm Z_{\delta m}^{\mathrm{GF}}$,
so the false-alarm probability stays close to $\alpha$ only if the systematics are
controlled well enough that $\lambda_{\delta m}\ll1$. This is a requirement not on the overall accuracy of the waveform
model but on its projection onto the beyond-vacuum direction, and it can be used in
future studies on the impact of waveform systematics in  beyond-vacuum GR effects.

\subsection{Relationship to the Savage-Dickey ratio}
\label{sec:SDsignificanceGF}
We have formulated the detection statistic in terms of a comparison of maximum likelihoods. However, in gravitational-wave astronomy it is more usual to carry out model comparison using Bayesian evidence ratios. Within the linear signal approximation we employed, these are actually equivalent. For the comparison of nested hypotheses of the form
\begin{equation}
    H_0: \; \thb = 0, \, \thv, \, \thc \qquad \mbox{vs} \qquad H_1: \; \thb, \, \thv , \, \thc \;,
\end{equation}
under the assumption that the priors on the common model parameters, $\thv$ and $\thc$, are consistent between the two hypotheses, the log evidence ratio (log Bayes factor) for $H_1$ over $H_0$ reduces to the Savage-Dickey ratio~\cite{dickey1971weighted, Cornish:2007if}
\begin{equation}
    \log(\mbox{BF}_{\rm SD}) = \log\left[p(\thb=0|H_1)\right] - \log\left[p(\thb=0|\mathbf{d},H_1)\right],
\end{equation}
which is simply the difference in the log-prior and log-posterior probability of the null hypothesis under the alternative hypothesis. If we assume that we place a flat prior on the beyond-vacuum parameters and that the allowed prior volume is large enough to contain the bulk of the posterior distribution, so that the integral may be approximated as an integral to infinity, we obtain
\begin{equation}
\log({\rm BF}_{\rm SD}) = S^{\rm GF}
  - \log\biggl[\frac{\Pi_{\rm b}\,\sqrt{\det\Fc{b}{cv}}}{(2\pi)^{N_{\rm b}/2}}\biggr] ,
\end{equation}
in which $\Pi_{\rm b}$ is the volume of the prior space for the beyond-vacuum
parameters, so that the argument of the logarithm is the ratio of the prior
volume to that of the posterior, $(2\pi)^{N_{\rm b}/2}(\det\Fc{b}{cv})^{-1/2}$.
The prior volume must be bigger than a few times
$(\det\Fc{b}{cv})^{-1/2}$ so that it completely contains the posterior. Making the volume even bigger reduces the Bayes factor. We conclude that the biggest log Bayes factor will be approximately $S^{\mathrm{GF}}-O(N_b)$. So, a threshold on $S^{\mathrm{GF}}$ computed from a $\chi^{2}_{N_{\mathrm b}}$ distribution is equivalent to setting a threshold on the log-evidence to be greater than a few. The fact that the choice of $\Pi_{\rm b}$ impacts the Bayes factor simply reflects the usual prior dependence of Bayesian model comparison. In the following we will not worry about this further, but will base our analysis on $S^{\mathrm{GF}}$ and related quantities, as this is the key quantity in both frequentist and Bayesian analyses.

\section{Post-processing strategy detection statistic}
\label{sec:StrategyPP}

We now want to obtain a similar detection statistic in the post-processing
strategy, where the data is
first fitted with vacuum templates
\begin{equation}
m^{\mathrm{PP}}(\thc,\thv)=\hc(\thc)+h(\thv,\bm 0) \, ,
\label{eq:mPP}
\end{equation}
and the target source of interest is
then reanalysed on the residual left after subtracting the nuisance sources. Here, we assume that the adopted post-processing strategy involves subtracting the best-fit nuisance sources and using the single residual obtained in this way for the subsequent reanalysis. Later we will consider an alternative strategy in which we also marginalise over the uncertainty in the initial parameter estimation.

The possible bias in $\thb$ comes in two steps. First, the vacuum fit biases the
recovered nuisance-source parameters, because those can absorb the unmodelled
beyond-vacuum effect. Second, those biased sources are subtracted from the data,
carrying the bias to the residual used for the beyond-vacuum inference in
post-processing analyses. We compute these two contributions below to obtain
$\Delta\thb_{\mathrm{PP}}$ and the corresponding detection statistic
$S^{\mathrm{PP}}$.

\subsection{First step: the vacuum global fit}
\label{sec:PPstep1}

Let $\thv_{\star}$ denote the vacuum parameters that best reproduce the true signal
of the target source of interest, that is, the minimiser over $\thv$ of
$\nip{\delta h}{\delta h}$ with
\begin{equation}
\delta h = h(\thv_0,\thb_0)-h(\thv_{\star},\bm 0)
\simeq\partial_{\thb}h\,\thb_0-\partial_{\thv}h\,(\thv_{\star}-\thv_0) \, .
\label{eq:deltah_LSA}
\end{equation}
Minimising then gives
$\thv_{\star}-\thv_0=\F{vv}^{-1}\F{vb}\,\thb_0$, and hence
\begin{equation}
\delta h=\Pjperp{\Vs}\bigl(\partial_{\thb}h\bigr)\,\thb_0 ,
\qquad
\nip{\delta h}{\partial_{\thv}h}=\bm 0 ,
\qquad
\nip{\delta h}{\delta h}=(\thb_0)^{T}\Fc{b}{v}\,\thb_0 \, .
\label{eq:deltah_thetab}
\end{equation}

We now expand the vacuum global fit around
$\bm\theta_{\star}=(\thc_0,\thv_{\star},\bm 0)$ and write $\Delta_{\star}$ for
the displacement each block has from that point at the maximum likelihood solution. The residual
is
\begin{equation}
r\simeq n+\delta m+\delta h
-\partial_{\thc}\hc\,\Delta_\star\thc-\partial_{\thv}h\,\Delta_\star \thv.
\label{eq:residual}
\end{equation}
The model error $\delta m$  enters the analysis
in the same way as $\delta h$, since both are pieces of data the vacuum templates
cannot represent. Minimising $\nip{r}{r}$ gives
\begin{equation}
\begin{pmatrix}\Delta_\star\thc\\\Delta_\star\thv\end{pmatrix}
=\begin{pmatrix}\F{cc}&\F{cv}\\ \F{vc}&\F{vv}\end{pmatrix}^{-1}
\begin{pmatrix}\nip{\partial_{\thc}\hc}{n+\delta m+\delta h}\\[3pt]
\nip{\partial_{\thv}h}{n+\delta m+\delta h}\end{pmatrix},
\label{eq:ML}
\end{equation}
which is similar to the bias formula~\eqref{eq:cvbias} applied to the first vacuum fit, with
the difference that the displacement is measured from $\bm\theta_{\star}$ rather
than from the true parameters. Substituting back into Eq.~\eqref{eq:residual}, we find that the
best-fit residual is
\begin{equation}
r_{\mathrm{bf}}=\Pjperp{\Vs\oplus\Cs}\bigl(n+\delta m+\delta h\bigr) ,
\label{eq:rbf}
\end{equation}
so the fit removes whatever lies in the vacuum tangent space, in agreement with the geometric intuition of
Sec.~\ref{sec:geometry}. Now, inverting Eq.~\eqref{eq:ML} blockwise using Eq.~\eqref{eq:InvBlock}
(with $\rm Z=v$, $\rm X=c$)
\begin{equation}
\Delta_\star\thc=\Fc{c}{v}^{-1}\nip{\Pjperp{\Vs}\bigl(\partial_{\thc}\hc\bigr)}{n+\delta m+\delta h} \, .
\label{eq:Deltac}
\end{equation}
Isolating the term corresponding to the unmodelled beyond-vacuum effect
\begin{equation}
\bigl(\Delta_\star\thc\bigr)_{\delta h}
=\Fc{c}{v}^{-1}\nip{\Pjperp{\Vs}\bigl(\partial_{\thc}\hc\bigr)}{\delta h}=\Fc{c}{v}^{-1}\Gamma_{\mathrm{cb}\mid\mathrm{v}}\,\thb_0 \, .
\label{eq:Deltac_dh}
\end{equation}
This is where post-processing loses relative to the global fit. The part of the
effect absorbed by the target source's own vacuum parameters,
$\thv_{\star}-\thv_0$, can be recovered in the second fit of the post-processing
stage, which frees $\thv$ again. But whatever correlates with the nuisance sources
($\Gamma_{\mathrm{cb}\mid\mathrm{v}}\neq\bm 0$) is absorbed into their parameters
in the first fit, and is subtracted from the data with them, never reaching the
second step of the reanalysis. Note that Eq.~\eqref{eq:Deltac_dh} is just the
multi-source (Cutler–Vallisneri) bias of Ref.~\cite{Antonelli:2021vwg}, with their generic waveform error applied
to the parametrised effect $\partial_{\thb}h \;\thb_0$. In the
language of Refs.~\cite{Yunes:2009ke,Vallisneri:2013rc}, it is the fundamental
bias that an unmodelled effect in the target induces in its neighbours. This bias
is stealthy whenever it exceeds their statistical uncertainty in the vacuum fit,
set by $\Fc{c}{v}^{-1}$, while the effect itself remains below the detection
threshold of the global fit, $(\thb_0)^{T}\Fc{b}{cv}\,\thb_0\lesssim 2S_{*}$
(Sec.~\ref{sec:significanceGF}).

\subsection{Second step: reanalysis of the residual}
\label{sec:PPstep2}

Subtracting the fitted, possibly biased nuisance sources from the data leaves the residual
\begin{equation}
\tilde r=d-\hc(\thc_0+\Delta_\star\thc)
=n+\delta m+h(\thv_0,\thb_0)+\Delta h ,
\qquad
\Delta h=-\partial_{\thc}\hc\,\Delta_\star\thc \, .
\label{eq:Deltah_sys}
\end{equation}
We now
fit $h(\thv,\thb)$ to $\tilde r$. This time, we expand around the true parameters
$(\thv_0,\thb_0)$ instead of $\bm\theta_{\star}$, so that what we obtain is
the bias $\Delta\thb_{\mathrm{PP}}$ on the beyond-vacuum parameters, which is directly
comparable with the $\Delta\thb_{\mathrm{GF}}$ of Sec.~\ref{sec:StrategyGF}.

Applying the bias formula~\eqref{eq:cvbias} with the two-block
rule (Eq.~\eqref{eq:InvBlock} with $\rm Z=v$, $\rm X=b$), in analogy with
Eq.~\eqref{eq:bGF},
\begin{equation}
\Delta\thb_{\mathrm{PP}}=\Fc{b}{v}^{-1}\bm Z^{\mathrm{PP}} ,
\qquad
\bm Z^{\mathrm{PP}}=
\nip{\Pjperp{\Vs}\bigl(\partial_{\thb}h\bigr)}{n+\delta m+\Delta h} \, .
\label{eq:bPP}
\end{equation}
Note that now $\Fc{b}{v}$ appears instead of $\Fc{b}{cv}$, because this step
marginalises over $\thv$ but $\thc$ has already been fixed at its biased
value and subtracted from the data. Also, Eq.~\eqref{eq:bPP} is not the sum of independent contributions because $\Delta h$ is itself built from $n$, $\delta m$ and $\delta h$ through
Eq.~\eqref{eq:Deltac}. Expanding it and separating into these 3 terms
\begin{equation}
\bm Z^{\mathrm{PP}}=\bm Z_n^{\mathrm{GF}}+\bm Z_{\delta m}^{\mathrm{GF}}
+\bm Z_{\delta h}^{\mathrm{PP}} ,
\qquad
\bm Z_{\delta h}^{\mathrm{PP}}
=-\Gamma_{\mathrm{cb}\mid\mathrm{v}}^{T}\Fc{c}{v}^{-1}
  \Gamma_{\mathrm{cb}\mid\mathrm{v}}\,\thb_0
=-\bigl(\Fc{b}{v}-\Fc{b}{cv}\bigr)\thb_0 ,
\label{eq:Zsplit_pp}
\end{equation}
where $\bm Z_{\bullet}^{\mathrm{GF}}$ are the global-fit overlaps of
Eq.~\eqref{eq:Zsplit_gf} and the last equality follows from
Eq.~\eqref{eq:cGamma}.
The last term, $Z_{\delta h}^{\mathrm{PP}}$, is precisely the part of the effect absorbed by the nuisance sources in the
first step. It is set by the difference between the two conditional Fisher
matrices, and vanishes when the nuisance sources do not correlate with the
beyond-vacuum direction of the target, so that $\Fc{cb}{v}=\mathbf{0}$ and
$\Fc{b}{v}=\Fc{b}{cv}$.

The corresponding statistic, the analogue of Eq.~\eqref{eq:Sgf_def}, is
\begin{equation}
S^{\mathrm{PP}}=
\tfrac12\bigl(\thb_0+\Delta\thb_{\mathrm{PP}}\bigr)^{T}\Fc{b}{v}
\bigl(\thb_0+\Delta\thb_{\mathrm{PP}}\bigr) \, .
\label{eq:maskcrit}
\end{equation}

By Eq.~\eqref{eq:Zsplit_pp} the beyond-vacuum term can be absorbed into the
prefactor,
$\thb_0+\Fc{b}{v}^{-1}\bm Z_{\delta h}^{\mathrm{PP}}
=\Fc{b}{v}^{-1}\Fc{b}{cv}\,\thb_0$, so that
\begin{equation}
(\thb_{\rm bf})^{\mathrm{PP}} = \thb_0+\Delta\thb_{\mathrm{PP}}
=\Fc{b}{v}^{-1}\bigl[\Fc{b}{cv}\,\thb_0
+\bm Z_{\delta m}^{\mathrm{GF}}+\bm Z_{n}^{\mathrm{GF}}\bigr] \, .
\label{eq:PPrecovered}
\end{equation}
Averaging $S^{\mathrm{PP}}$ over noise realisations, the term linear in
$\bm Z_n^{\mathrm{GF}}$ vanishes as before, and the quadratic one gives
\begin{equation}
\tfrac12\bigl\langle(\bm Z_n^{\mathrm{GF}})^{T}\Fc{b}{v}^{-1}
\bm Z_n^{\mathrm{GF}}\bigr\rangle
=\tfrac12\Tr\bigl[\Fc{b}{v}^{-1}\,\mathrm{Cov}(\bm Z_n^{\mathrm{GF}})\bigr]
=\tfrac12\Tr\bigl[\Fc{b}{v}^{-1}\Fc{b}{cv}\bigr] \leq \frac{1}{2}N_{\rm b}\, .
\label{eq:PPfloor}
\end{equation}
The last inequality follows from $\Fc{b}{cv}\preceq\Fc{b}{v}$ and that Fisher matrices are positive semi-definite by construction (we show this in Appendix~\ref{app:lemma}).  The expectation of the post-processing statistic in the null case, when no extra effect is present in the data, is
therefore smaller than that of the global fit. This implies that the recovered parameters fluctuate
less across noise realisations than the quoted uncertainty in post-processing. In this strategy, the second step assumes that only the target source is present in the data and therefore all noise degrees of freedom can impact parameter estimation. In reality, the noise degrees of freedom that also correlate with the nuisance sources have been removed in the first stage of the fit, and therefore the residual uncertainty is less than the model assumes. Therefore, the posterior uncertainty overestimates the true fluctuations, unless $\Fc{b}{v} = \Fc{b}{cv}$, i.e. unless the nuisance sources do not correlate with the beyond-vacuum direction
of the target source.

Averaging Eq.~\eqref{eq:maskcrit} over noise with Eq.~\eqref{eq:PPrecovered}, as
done for the global fit in Eq.~\eqref{eq:maskgf_u},
\begin{equation}
\langle S^{\mathrm{PP}}\rangle
=\tfrac12\bigl[(\thb_0)^{T}\Fc{b}{cv}\Fc{b}{v}^{-1}\Fc{b}{cv}\thb_0
+\Tr\bigl(\Fc{b}{v}^{-1}\Fc{b}{cv}\bigr)\bigr]
+(\thb_0)^{T}\Fc{b}{cv}\Fc{b}{v}^{-1}\bm Z_{\delta m}^{\mathrm{GF}}
+\tfrac12\,(\bm Z_{\delta m}^{\mathrm{GF}})^{T}\Fc{b}{v}^{-1}
 \bm Z_{\delta m}^{\mathrm{GF}} \, ,
\label{eq:maskpp_u}
\end{equation}
where the different terms have a similar interpretation and can also be combined into a perfect
square at the \emph{same} effective parameter $\thb_{\mathrm{eff}}$ of
Eq.~\eqref{eq:thetaeff}
\begin{equation}
\langle S^{\mathrm{PP}}\rangle
=\tfrac12\bigl[(\thb_{\mathrm{eff}})^{T}\Fc{b}{cv}\Fc{b}{v}^{-1}\Fc{b}{cv}
\thb_{\mathrm{eff}}+\Tr\bigl(\Fc{b}{v}^{-1}\Fc{b}{cv}\bigr)\bigr] \, .
\label{eq:maskpp_eff}
\end{equation}
Comparing Eq.~\eqref{eq:PPrecovered} with
Eq.~\eqref{eq:bGF}, we see that
\begin{equation}
\thb_0+\Delta\thb_{\mathrm{PP}}
=\Fc{b}{v}^{-1}\Fc{b}{cv}\,\bigl(\thb_0+\Delta\thb_{\mathrm{GF}}\bigr) \, .
\label{eq:PPmap}
\end{equation}
Post-processing
does not recover $\thb_{\mathrm{eff}}$ but
\begin{equation}
\bigl\langle(\thb_{\rm bf})^{\mathrm{PP}}\bigr\rangle
=\Fc{b}{v}^{-1}\Fc{b}{cv}\,\thb_{\mathrm{eff}} \, .
\end{equation}
So, the best fit estimate in post-processing is biased by the fraction of the
effect the nuisance sources absorbed. As we show below, this bias is always towards vacuum.

\subsection{Detection significance}
\label{sec:significancePP}

The best-fit residuals obtained by fitting $h(\thv,\thb)$ and $h(\thv,\bm 0)$ to
the residual $\tilde r$ of Eq.~\eqref{eq:Deltah_sys} differ by
\begin{equation}
\nip{\tilde r_{\rm vac}}{\tilde r_{\rm vac}}
-\nip{\tilde r_{\rm bf}}{\tilde r_{\rm bf}}
=\bigl((\thb_{\rm bf})^{\mathrm{PP}}\bigr)^{T}\Fc{b}{v}\,(\thb_{\rm bf})^{\mathrm{PP}}
=2S^{\mathrm{PP}} \, ,
\label{eq:LLRpp}
\end{equation}
so $S^{\mathrm{PP}}$ is the right statistic to use for claiming
a deviation from GR in the post-processing strategy. It also gives the Bayes factor for the presence of a beyond-vacuum GR effect, as discussed in Sec.~\ref{sec:SDsignificanceGF}. The distribution of $S^{\mathrm{PP}}$ is not a $\chi^{2}$, because the recovered value $\thb_{\rm bf}$ is not
being compared against its covariance. Applying
$\mathrm{Cov}[Ax]=A\,\mathrm{Cov}[x]\,A^{T}$ to Eq.~\eqref{eq:PPrecovered}
\begin{equation}
\mathrm{Cov}\bigl[(\thb_{\rm bf})^{\mathrm{PP}}\bigr]
=\Fc{b}{v}^{-1}\,\Fc{b}{cv}\,\Fc{b}{v}^{-1}
\;\neq\;\Fc{b}{v}^{-1} \, .
\label{eq:covPP}
\end{equation}
For a Gaussian distribution $x\sim\mathcal{N}(\mu,\Sigma)$, the quadratic form $x^{T}Wx$ is a
$\chi^{2}$ only when $W=\Sigma^{-1}$. In general, it follows a weighted sum of
independent non-central $\chi^{2}_{1}$, with weights given by the eigenvalues of
$W\Sigma$. Here, $W=\Fc{b}{v}$ and $\Sigma$ is Eq.~\eqref{eq:covPP}, so that
$W\Sigma=\Fc{b}{cv}\Fc{b}{v}^{-1}$ and
\begin{equation}
2S^{\mathrm{PP}}\ \sim\ \sum_{i=1}^{N_{\mathrm b}} w_i\,\chi^{2}_{1}(\nu_i) ,
\quad
\{w_i\}=\mathrm{eig}\bigl(\Fc{b}{v}^{-1}\Fc{b}{cv}\bigr)\in [0,1] ,
\quad
\nu_i=\bigl(u_i^{T}\,\Fc{b}{cv}^{1/2}\,\thb_{\mathrm{eff}}\bigr)^{2} ,
\label{eq:chi2pp}
\end{equation}
with $w_i$ and $u_i$ being, respectively, the eigenvalues and orthonormal
eigenvectors of the symmetric matrix
$\Fc{b}{cv}^{1/2}\Fc{b}{v}^{-1}\Fc{b}{cv}^{1/2}$. We show that $w_i$ are real, positive and no greater than
unity in
Appendix~\ref{app:lemma}. The non-centrality parameters are given by
\begin{equation}
\sum_i \nu_i=\lambda^{\mathrm{GF}} ,
\qquad
\sum_i w_i\,\nu_i=\lambda^{\mathrm{PP}} \, .
\label{eq:sumrules}
\end{equation}
Here, $\lambda^{\rm PP}\equiv(\thb_{\mathrm{eff}})^{T}\Fc{b}{cv}\Fc{b}{v}^{-1}\Fc{b}{cv}\,\thb_{\mathrm{eff}}$
is the signal term of Eq.~\eqref{eq:maskpp_eff}.
In the global fit strategy, $w_i=1$ and Eq.~\eqref{eq:chi2pp} becomes the $\chi^{2}_{N_{\mathrm b}}$ of
Eq.~\eqref{eq:chi2}. Also, note that for a single beyond-vacuum parameter there is only one
weight, $w=\Fc{b}{cv}/\Fc{b}{v}$, and the two statistics are then related by a
constant, $2S^{\mathrm{PP}}=\left( \Fc{b}{cv}/\Fc{b}{v} \right)\,2S^{\mathrm{GF}}$, for every noise realisation.

\subsection{Accounting for the uncertainty of the subtracted nuisance sources}
\label{sec:marginalised}

The two-step strategy above assumes that a single residual will be used in post-processing by subtracting
the nuisance sources using
the point estimate
$\thc_0+\Delta_\star\thc$ of the first fit. But the first fit also provides
the uncertainty of that estimate, and so an alternative would be to
propagate this uncertainty into the second stage. Instead of subtracting the best fit, one can draw samples
$\thc_s$ from the posterior of the vacuum global fit, subtract each, repeat
the second step of the post-processing analysis, and \textit{average} the
resulting posteriors. The additional cost compared to the simple
post-processing strategy is then one low-dimensional refit per draw.

At linear-signal order, the effect of averaging can be inferred following Sec.~\ref{sec:PPstep2}. The nuisance parameters of a draw $s$ are
$\thc_s=\thc_0+\Delta_\star\thc+\delta\thc_s$, where $\delta\thc_s$ is the
fluctuation of the subtracted configuration around the first-step best fit.
Its covariance is the $\thc$ block of the inverse of the first-step Fisher
matrix in Eq.~\eqref{eq:ML}, which by Eq.~\eqref{eq:InvBlock} is
$\Fc{c}{v}^{-1}$. Subtracting $\hc(\thc_s)$ instead of
$\hc(\thc_0+\Delta_\star\thc)$ adds $-\partial_{\thc}\hc\,\delta\thc_s$ to
the residual $\tilde r$ of Eq.~\eqref{eq:Deltah_sys}, which in turn adds
$-\Gamma_{\mathrm{cb}\mid\mathrm{v}}^{T}\,\delta\thc_s$ to
$\bm Z^{\mathrm{PP}}$ in Eq.~\eqref{eq:bPP}, and displaces the
recovered value of Eq.~\eqref{eq:PPrecovered} by
\begin{equation}
\delta\thb_s
=-\,\Fc{b}{v}^{-1}\,\Gamma_{\mathrm{cb}\mid\mathrm{v}}^{T}\,\delta\thc_s \, .
\label{eq:dphib}
\end{equation}
Since $\delta\thc_s$ is drawn from a Gaussian, $\delta\thb_s$ follows a zero-mean Gaussian with covariance
\begin{equation}
\mathrm{Cov}\bigl[\delta\thb\bigr]
=\Fc{b}{v}^{-1}\,
\Gamma_{\mathrm{cb}\mid\mathrm{v}}^{T}\Fc{c}{v}^{-1}
\Gamma_{\mathrm{cb}\mid\mathrm{v}}\,
\Fc{b}{v}^{-1}
=\Fc{b}{v}^{-1}\bigl(\Fc{b}{v}-\Fc{b}{cv}\bigr)\Fc{b}{v}^{-1} \, ,
\label{eq:covdphi}
\end{equation}
where the last equality follows from Eq.~\eqref{eq:cGamma}.

Each per-draw posterior is the Gaussian of Sec.~\ref{sec:PPstep2} displaced
by $\delta\thb_s$, which is itself a zero-mean Gaussian. Averaging them then leads to a Gaussian posterior with the same mean as simple post-processing, Eq.~\eqref{eq:PPrecovered}, and with covariance the sum of that of simple post-processing and
that of the shifts $\delta\thb$
\begin{equation}
\mathrm{Cov}^{\PPb}\bigl[\thb\bigr]
=\Fc{b}{v}^{-1}+\mathrm{Cov}\bigl[\delta\thb\bigr]
=\Fc{b}{v}^{-1}\bigl(2\,\Fc{b}{v}-\Fc{b}{cv}\bigr)\Fc{b}{v}^{-1} \, ,
\label{eq:margvar}
\end{equation}
where the overbar denotes the average over the posterior draws.

The detection statistic to consider is now the distance from vacuum of the
averaged posterior, measured with this new covariance
\begin{equation}
S^{\PPb}=\tfrac12\bigl((\thb_{\rm bf})^{\mathrm{PP}}\bigr)^{T}
\bigl(\mathrm{Cov}^{\PPb}\bigl[\thb\bigr]\bigr)^{-1}
(\thb_{\rm bf})^{\mathrm{PP}} \, ,
\label{eq:SPPb}
\end{equation}
which is the analogue of Eq.~\eqref{eq:maskcrit} for simple post-processing. Substituting Eq.~\eqref{eq:PPrecovered}
and taking the noise average as for Eq.~\eqref{eq:maskpp_eff}
\begin{equation}
\bigl\langle S^{\PPb}\bigr\rangle
=\tfrac12\Bigl[(\thb_{\mathrm{eff}})^{T}\,\Fc{b}{cv}
\bigl(2\,\Fc{b}{v}-\Fc{b}{cv}\bigr)^{-1}\Fc{b}{cv}\,\thb_{\mathrm{eff}}
+\Tr\Bigl(\bigl(2\,\Fc{b}{v}-\Fc{b}{cv}\bigr)^{-1}\Fc{b}{cv}\Bigr)\Bigr] \, .
\label{eq:Smarg}
\end{equation}
In the next section we compare the detection statistics under these different strategies.

\section{The Masking Ratio}
\label{sec:comparison}

We start by quantifying what simple post-processing
loses with respect to the full global fit. The difference between their detection statistics is
\begin{equation}
\Delta S=\langle S^{\mathrm{GF}}\rangle-\langle S^{\mathrm{PP}}\rangle
=\tfrac12\,(\thb_{\mathrm{eff}})^{T}
   \bigl[\Fc{b}{cv}-\Fc{b}{cv}\Fc{b}{v}^{-1}\Fc{b}{cv}\bigr]\thb_{\mathrm{eff}}
+\tfrac12\Bigl[N_{\mathrm b}-\Tr\bigl(\Fc{b}{v}^{-1}\Fc{b}{cv}\bigr)\Bigr]
\;\ge\;0 \, ,
\label{eq:DeltaS}
\end{equation}
and both terms are non-negative, as we show in Appendix~\ref{app:lemma}. So, on
average and in the linear-signal approximation, post-processing can only
mask a beyond-vacuum GR effect. When no such effect is present, the
bound it sets on a deviation from GR is spuriously tight. We consider both cases below, and then combine them into a single measure in Eq.~\eqref{eq:Rmin}.

It may seem counter-intuitive that the post-processing strategy is both too conservative, potentially missing deviations that are present, and too optimistic, setting bounds that are too tight. This is because it gets both the mean and the width of the posterior wrong. The width is determined by the number of degrees of freedom being fitted and this is fewer in the post-processing strategy as the nuisance sources are not included. The mean, however, is shifted by an amount that is determined from the correlations between the nuisance and target sources. The post-processing strategy assumes all degrees of freedom in the target source are available in the fit, while some of these are removed because they correlate with the nuisance sources. Thus, the mean is shifted closer to zero than the post-processing model assumes is possible. Overall, post-processing produces posteriors that are too narrow, but are also closer to zero and fluctuating less than the post-processing model expects. The result is that we are both over-confident in upper-limits on deviations from vacuum-GR, and, when the detection threshold is set from the
post-processing likelihood, less likely to detect a deviation that is present.

For a sufficiently strong beyond-vacuum effect, the significance of a global-fit
detection approaches $\sqrt{\lambda^{\mathrm{GF}}}$, the square root of the
non-centrality of its $\chi^{2}$ distribution
(Sec.~\ref{sec:significanceGF}). We then define the masking ratio for any number of beyond-vacuum
parameters, generalising Eq.~\eqref{eq:R_geometric}, as
\begin{equation}
\mathcal{R}\;=\;
\sqrt{\frac{\lambda^{\mathrm{PP}}}{\lambda^{\mathrm{GF}}}}
=\left[
\frac{(\thb_{\mathrm{eff}})^{T}\Fc{b}{cv}\Fc{b}{v}^{-1}\Fc{b}{cv}\,\thb_{\mathrm{eff}}}
     {(\thb_{\mathrm{eff}})^{T}\Fc{b}{cv}\,\thb_{\mathrm{eff}}}
\right]^{1/2}\;\in\;[0,1] \, .
\label{eq:Rdef}
\end{equation}
A deviation that the global fit would have detected at $n_\sigma$ is then
reported by post-processing at $\mathcal{R}\,n_\sigma$, when its
significance is computed, as usual, from the target-only likelihood.

When no effect is present, $\mathcal{R}$ can still be computed, since it is
independent of the amplitude of the effect. However, its interpretation
changes, as it is now related to the two possible failure modes of a null test.
The first is that we think a deviation is present when it is not. Fully
quantifying this requires a detection algorithm, but we assume for simplicity
that it is based on $S^{\rm PP}$. A threshold is placed on $S^{\rm PP}$, and a value above it is deemed a detection. Under the post-processing model
assumptions, $2S^{\rm PP}$ follows a $\chi^2_{N_{\rm b}}(0)$ distribution, which
would be used to set the detection threshold. In reality, it follows the
weighted sum of $\chi^2_1(0)$ distributions of Eq.~\eqref{eq:chi2pp}, and we
define the \emph{null ratio} as the ratio of its actual mean to the assumed one
\begin{equation}
\mathcal{R}_{0}\;=\;
\frac{\Tr\bigl(\Fc{b}{v}^{-1}\Fc{b}{cv}\bigr)}{N_{\mathrm b}}
\;\in\;[0,1] \, ,
\label{eq:R0def}
\end{equation}
so the factor by which the null expectation of $2S^{\mathrm{PP}}$ is below the
$\chi^{2}_{N_{\mathrm b}}$ on which the detection statistic would be based.

Since $\mathcal{R}_0\le1$, post-processing is on average less likely than the
global fit to claim a deviation that is not there, and, as we now show, this
holds at every significance level. In terms of the eigenvectors $u_i$ of
Eq.~\eqref{eq:chi2pp}, the two statistics are $2S^{\rm GF}=\sum_i x_i^2$ and
$2S^{\rm PP}=\sum_i w_i\,x_i^2$, with the same
$x_i=u_i^T\Fc{b}{cv}^{1/2}(\thb_{\rm bf})^{\rm GF}$ in both, because by
Eq.~\eqref{eq:PPrecovered} the two best fits are related by
$(\thb_{\rm bf})^{\rm PP}=\Fc{b}{v}^{-1}\Fc{b}{cv}\,(\thb_{\rm bf})^{\rm GF}$.
Since $w_i\le1$, $S^{\rm PP}\le S^{\rm GF}$ in
every noise realisation (for $N_{\rm b}=1$ they are simply proportional, as we show in Eq.~\eqref{eq:sig_loss} below). In the null case, the $x_i$ are independent
standard Gaussians, so $2S^{\rm GF}$ follows exactly the $\chi^2_{N_{\rm b}}(0)$
from which we set the detection threshold $2S_*$ in  post-processing. Hence
$P(2S^{\rm PP}>2S_*)\le P(2S^{\rm GF}>2S_*)=\alpha$ for any threshold, so
post-processing is a conservative strategy in the sense that it never
rejects vacuum GR more often than the nominal false-alarm probability.

The second failure mode is that, having found no evidence for a deviation, we place an upper
limit that is too high or too low. Again, exactly what this means depends on the
scenario, since there can be several beyond-vacuum parameters, constrained
individually or jointly, and the upper limit is really a statement about the
full beyond-vacuum posterior. The simplest thing to do is to compare the widths
of the global-fit and post-processing posteriors, and this again yields
$\mathcal{R}_0$. Now, its numerator can be interpreted as the expected squared
significance, measured with the global-fit posterior, of a point drawn at random
from the post-processing posterior, if both were centred at the origin. A value
$\mathcal{R}_0<1$ then implies that the post-processing posterior is more
compact than the global-fit one, and hence yields tighter upper limits. This
simple comparison neglects that the mean is shifted as well, by an amount
inconsistent with the posterior width, which also affects the upper limit. Inspired by the previous interpretation of $\mathcal{R}_0$, one way to compute a quantity that accounts for this is to consider the
distance at which samples randomly drawn from the post-processing posterior lie
from the centre of the global-fit posterior, in units of its width, now allowing
for the fluctuations in the mean as well as the differences in the width. In the null case, the global-fit
posterior is $\mathcal{N}(\Delta\thb_{\rm GF},\Fc{b}{cv}^{-1})$ and the
post-processing one is $\mathcal{N}(\Delta\thb_{\rm PP},\Fc{b}{v}^{-1})$, so a sample of the latter is $\Delta\thb_{\rm PP}+\Delta\thb_{\rm s}$, with
$\Delta\thb_{\rm s}\sim\mathcal{N}(0,\Fc{b}{v}^{-1})$, and the relevant
quantity is
\begin{equation}
    \bigl(\Delta\thb_{\rm PP}+\Delta\thb_{\rm s}-\Delta\thb_{\rm GF}\bigr)^T \Fc{b}{cv}
    \bigl(\Delta\thb_{\rm PP}+\Delta\thb_{\rm s}-\Delta\thb_{\rm GF}\bigr) .
    \label{eq:ULdist}
\end{equation}
For samples of the global-fit posterior itself, its expectation is $N_{\rm b}$.
If it is smaller, the post-processing posterior is too concentrated and any
upper limit is overly tight, and if it is larger than $N_{\rm b}$, upper limits are overly
conservative. Taking the expectation over $\Delta\thb_{\rm s}$ and over noise
realisations, and noting that
$\Delta\thb_{\rm PP}=\Fc{b}{v}^{-1}\Fc{b}{cv}\,\Delta\thb_{\rm GF}$ by
Eq.~\eqref{eq:PPrecovered}, we find
\begin{align}
    \mathcal{R}_{0,{\rm UL}}\;&=\;\frac{\bigl\langle\bigl(\Delta\thb_{\rm PP}+\Delta\thb_{\rm s}-\Delta\thb_{\rm GF}\bigr)^T \Fc{b}{cv} \bigl(\Delta\thb_{\rm PP}+\Delta\thb_{\rm s}-\Delta\thb_{\rm GF}\bigr)\bigr\rangle}{N_{\rm b}} \nonumber \\
    &= \frac{1}{N_{\rm b}}\Tr\Bigl[\bigl(\Fc{b}{v}^{-1}\Fc{b}{cv}\bigr)^2-\Fc{b}{v}^{-1}\Fc{b}{cv}+\mathbb{I}_{N_{\rm b}}\Bigr] \nonumber \\
    &= \frac{1}{N_{\rm b}}\sum_i \bigl[(1-w_i)^2+w_i\bigr] \;\in\;\bigl[\tfrac34,1\bigr] ,
\label{eq:RzUL}
\end{align}
where the first term comes from the shift of the mean,
$\Delta\thb_{\rm PP}-\Delta\thb_{\rm GF}$, and the second from the width,
$\Delta\thb_{\rm s}$. The upper limits obtained from post-processing are
therefore always overly tight. Additionally, since $\mathcal{R}_{0,{\rm UL}}-\mathcal{R}_0=\frac{1}{N_{\rm b}}\sum_i(1-w_i)^2\ge0$, we conclude that placing a criterion that the null ratio is sufficiently close to $1$ ensure that neither of the two failure modes of non-detections occurs. We will henceforth use only $\mathcal{R}_0$ to assess the potential of wrong inference through post-processing in the null case.

Note that the null ratio and the masking ratio are not independent, as they both
involve averages of the weights
\begin{equation}
\mathcal{R}^{2}=\frac{\sum_i w_i\,\nu_i}{\sum_i \nu_i} ,
\qquad
\mathcal{R}_{0}=\frac{1}{N_{\mathrm b}}\sum_i w_i \, .
\label{eq:twomeans}
\end{equation}
The weights depend only on the waveform models of the target and nuisance
sources, through the Fisher matrices at the expansion point, and not on the
noise realisation. Therefore, they, and $\mathcal{R}_{0}$ with them, can be
computed before anything is known about the effect. The masking ratio does not
depend on the amplitude of the effect either, since it is invariant under
$\thb_{\mathrm{eff}}\to c\,\thb_{\mathrm{eff}}$. For $N_{\mathrm b}>1$, however,
it does depend on the direction in the beyond-vacuum subspace along which the
effect is excited. Even then, it is bounded by the smallest and largest weights
\begin{equation}
\sqrt{w_{\mathrm{min}}}\;\le\;\mathcal{R}\;\le\;\sqrt{w_{\mathrm{max}}} \, ,
\label{eq:Rbracket}
\end{equation}
as we also show in Appendix~\ref{app:lemma}. Since $\mathcal{R}_0$ is the mean
of the weights, it lies in the same range, $w_{\min}\le\mathcal{R}_0\le w_{\max}$.

The masking and null ratios can be interpreted as the two limits of a single quantity, which we
call the \emph{survival ratio}, as it measures the fraction of the expected
global-fit statistic that survives post-processing, $1-\Delta S/\langle S^{\rm GF}\rangle$ with $\Delta S$ from Eq.~\eqref{eq:DeltaS}. By Eq.~\eqref{eq:chi2pp}
\begin{equation}
\mathcal{R}_{\lambda}^{2}
\equiv\frac{\langle 2S^{\rm PP}\rangle}{\langle 2S^{\rm GF}\rangle}
=\frac{\sum_i w_i\,(\nu_i+1)}{\sum_i(\nu_i+1)}
=\frac{\lambda^{\rm GF}\,\mathcal{R}^{2}+N_{\rm b}\,\mathcal{R}_{0}}
      {\lambda^{\rm GF}+N_{\rm b}} .
\label{eq:Rlambda}
\end{equation}
This quantity tends to $\mathcal{R}_0$ when no effect is present and to
$\mathcal{R}^2$ for a strong one, $\lambda^{\rm GF}\gg N_{\rm b}$. Unlike the
other two ratios, the survival ratio depends on the amplitude of the effect
through $\lambda^{\rm GF}$, which is not known before the analysis. However,
since it is a weighted average of $\mathcal{R}^2$ and $\mathcal{R}_0$, it always
lies between them, so that
\begin{equation}
\mathcal{R}_\lambda\;\ge\;\min\bigl(\mathcal{R},\sqrt{\mathcal{R}_0}\bigr)
\label{eq:Rmin}
\end{equation}
for an effect of any size. We therefore use the smaller of $\mathcal{R}$ and
$\sqrt{\mathcal{R}_0}$ as a single, conservative measure of how much
post-processing loses. When a specific beyond-vacuum model is tested,
$\mathcal{R}$ is evaluated along the direction of $\thb_{\rm eff}$ that it
predicts. Otherwise, the worst case over all directions is $\sqrt{w_{\min}}$,
which bounds $\mathcal{R}$, $\sqrt{\mathcal{R}_0}$ and $\mathcal{R}_\lambda$ from below, and is then the natural number with
which to select the sources to refit.

For a single beyond-vacuum GR parameter there is only one weight, and all three
ratios coincide
\begin{equation}
\mathcal{R}^{2}=\mathcal{R}_{0}=\mathcal{R}_\lambda^{2}=w=\frac{\Fc{b}{cv}}{\Fc{b}{v}} \, ,
\label{eq:Rscalar}
\end{equation}
independently of the tested value (a consequence of the linear-signal
approximation). A single number therefore describes both detections and null
tests, and the detection significances of the two strategies are related in
every noise realisation
\begin{equation}
\sqrt{2S^{\mathrm{PP}}}=\mathcal{R}\,\sqrt{2S^{\mathrm{GF}}} \, .
\label{eq:sig_loss}
\end{equation}
In post-processing, the best-fit value of the beyond-vacuum GR parameter shrinks
by $\mathcal{R}^{2}$, while the quoted uncertainty $\Fc{b}{v}^{-1/2}$ is tighter
than the global fit's $\Fc{b}{cv}^{-1/2}$ by only $\mathcal{R}$, so the error
bar on $\theta^{\rm b}$ shrinks at half the rate at which the nuisance sources
absorb the effect. This is the geometric result of
Eq.~\eqref{eq:R_geometric}, $\mathcal{R}=\sin\alpha$ (Fig.~\ref{fig:geometry}).
For $N_{\mathrm b}>1$, the weights $w_i$ of Eq.~\eqref{eq:chi2pp} are the
squared sines of the principal angles between the span of
$\Pjperp{\Vs}\,\partial_{\thb}m$ and $\Pjperp{\Vs}\Cs$. Although this is the simplest scenario, the case with
$N_{\rm b}=1$ covers many physical setups of interest~\cite{Speri:2024qak, Lara:2025kzj, Speri:2026ade, Fantoccoli:2026idl},
in which the extra effect enters the waveform at leading order through a single
dominant term whose amplitude is the only new parameter. This is also the case
for parametrised deviations from GR~\cite{Yunes:2009ke, Mehta:2022pcn}, in which
the deviation is expanded in a fixed basis (e.g. in weak-field post-Newtonian
order) and its coefficients are fitted to the data. These are usually
constrained one order at a time, so $N_{\rm b}=1$.

We can also compare the detection significance of the two post-processing
strategies (simple \textit{vs} averaged). They share the same point estimate
for the beyond-vacuum parameters, $(\thb_{\rm bf})^{\mathrm{PP}}$, and hence
its sampling covariance, Eq.~\eqref{eq:covPP}. The only difference is the
covariance against which it is compared in the detection statistic. In simple post-processing it is $\Fc{b}{v}$
in Eq.~\eqref{eq:LLRpp}, and in the average one it is
$\bigl(\mathrm{Cov}^{\PPb}[\thb]\bigr)^{-1}
=\Fc{b}{v}\bigl(2\,\II-\Fc{b}{v}^{-1}\Fc{b}{cv}\bigr)^{-1}$ in
Eq.~\eqref{eq:SPPb}. The latter is a function of the same matrix
$\Fc{b}{v}^{-1}\Fc{b}{cv}$ whose eigenvalues define the weights $w_i$ in
Eq.~\eqref{eq:chi2pp}, so in that decomposition only the weights change
\begin{equation}
2S^{\PPb}\ \sim\ \sum_{i=1}^{N_{\mathrm b}} \bar w_i\,\chi^{2}_{1}(\nu_i) ,
\qquad
\frac{w_i}{2}\;\le\;\bar w_i= \frac{w_i}{2-w_i}\;\le\;w_i \, ,
\label{eq:chi2pool}
\end{equation}
with the same non-centrality parameters $\nu_i$. The masking and null ratios of the
averaged strategy then follow from Eq.~\eqref{eq:twomeans} with
$w_i\to\bar w_i$
\begin{equation}
\bar{\mathcal{R}}^{2}=\frac{\sum_i \bar w_i\,\nu_i}{\sum_i \nu_i}
\;\le\;\mathcal{R}^{2} ,
\qquad
\bar{\mathcal{R}}_{0}=\frac{1}{N_{\mathrm b}}\sum_i \bar w_i
\;\le\;\mathcal{R}_{0} \, ,
\label{eq:Rpool}
\end{equation}
with the same bracketing by the smallest and largest weights as in
Eq.~\eqref{eq:Rbracket}, and the survival ratio of Eq.~\eqref{eq:Rlambda}
decreases likewise. Since $w_i/2 \le \bar w_i \le w_i$, the significance of a detection decreases, but by at most a factor of $\sqrt{2}$ and by much less when any $w_i \approx 1 $. This happens because averaging broadens the posterior, while not shifting the peak, bringing the distribution statistically closer to zero.  In simple post-processing, the shrinking of the recovered value
by $\mathcal{R}^2$ is partly compensated by the tighter uncertainty, as noted
after Eq.~\eqref{eq:sig_loss}, and averaging undoes part of this compensation
by correcting the width but not the mean. This suggests that, for a detection,
it is better to subtract only the best fit, which is also computationally
cheaper. However, this might be less robust to mismodelling, and it would be the
wrong strategy for setting an upper limit, as we now argue.

In the null case, the expectation of the statistic decreases likewise, to
$\bar{\mathcal{R}}_0\le\mathcal{R}_0$. This means we would be less likely to
falsely identify a beyond-vacuum GR deviation when none is present. However,
the posterior also widens, becoming more consistent with the global-fit one,
since
$\tfrac{1}{N_{\rm b}}\Tr\bigl(\Fc{b}{v}^{-1}\Fc{b}{cv}\bigr)\le
\tfrac{1}{N_{\rm b}}\Tr\bigl[\bigl(2\,\mathbb{I}_{N_{\rm b}}-\Fc{b}{v}^{-1}\Fc{b}{cv}\bigr)\Fc{b}{v}^{-1}\Fc{b}{cv}\bigr]\le1$,
where the middle term is the averaged covariance of Eq.~\eqref{eq:margvar} in
units of the global-fit one. We can also evaluate $\mathcal{R}_{0,{\rm UL}}$, as
defined in Eq.~\eqref{eq:RzUL}, for this case. The mean is unchanged, but the
covariance of $\Delta\thb_{\rm s}$ is now
$\bigl(2\,\mathbb{I}_{N_{\rm b}}-\Fc{b}{v}^{-1}\Fc{b}{cv}\bigr)\Fc{b}{v}^{-1}$,
Eq.~\eqref{eq:margvar}, which modifies $\mathcal{R}_{0,{\rm UL}}$ to
\begin{align}
\bar{\mathcal{R}}_{0,{\rm UL}}
&=\frac{1}{N_{\rm b}}\Tr\Bigl[\bigl(\mathbb{I}_{N_{\rm b}}-\Fc{b}{v}^{-1}\Fc{b}{cv}\bigr)^2
+\bigl(2\,\mathbb{I}_{N_{\rm b}}-\Fc{b}{v}^{-1}\Fc{b}{cv}\bigr)\Fc{b}{v}^{-1}\Fc{b}{cv}\Bigr] \nonumber\\
&=\frac{1}{N_{\rm b}}\sum_i\bigl[(1-w_i)^2+w_i(2-w_i)\bigr]=1 ,
\label{eq:R0ULbar}
\end{align}
where only the second, width term has changed. By this metric, samples of the
averaged posterior lie, on average, as far from the global-fit centre as samples
of the global fit itself. The metric equals one for any weights, so it cannot by
itself show that the bounds are exact, and for $N_{\rm b}=1$ the averaged
posterior is still narrower than the global-fit one by a factor
$\mathcal{R}\sqrt{2-\mathcal{R}^2}$, Eq.~\eqref{eq:margscalar}. Averaging
nevertheless brings the bounds closer to those of the global fit, which suggests
that it is the better post-processing
strategy for setting upper limits when no deviation is detected. This is
intuitively sensible, as when the correction is absent, the vacuum model of the
first fit is correct, and its full posterior preserves the information about
the correlations between the target and the nuisance sources, so we get close to
the true upper limit in post-processing. Note that, unlike for simple
post-processing, $\bar{\mathcal{R}}_0$ is then no guide to the upper limits since it
moves away from one, while the upper limits move closer to those of the global fit.

For a single beyond-vacuum parameter there is again only one weight,
$\bar w=w/(2-w)$ with $w=\mathcal{R}^{2}$, so all the ratios of the averaged
strategy coincide as well, and
\begin{equation}
\sigma^{2}_{\PPb}=\bigl(2-\mathcal{R}^{2}\bigr)\,\Fc{b}{v}^{-1} ,
\qquad
\bar{\mathcal{R}}^{2}=\bar{\mathcal{R}}_{0}=\bar{\mathcal{R}}_\lambda^{2}
=\frac{\mathcal{R}^{2}}{2-\mathcal{R}^{2}} ,
\qquad
\sqrt{2S^{\PPb}}=\frac{\sqrt{2S^{\mathrm{PP}}}}{\sqrt{2-\mathcal{R}^{2}}}
=\bar{\mathcal{R}}\,\sqrt{2S^{\mathrm{GF}}} \, ,
\label{eq:margscalar}
\end{equation}
where the last two equalities hold in every noise realisation, as in
Eq.~\eqref{eq:sig_loss}. The recovered value is still
$\mathcal{R}^{2}\thb_{\mathrm{eff}}$, with a posterior wider than in simple
post-processing by $\sqrt{2-\mathcal{R}^{2}}$ and narrower than the
global-fit one by $\mathcal{R}\sqrt{2-\mathcal{R}^{2}}$.

In summary, when a deviation is present, the averaging strategy gives results
that are more conservative than those of simple post-processing, and further
from the global fit, while when no deviation is present it gives upper limits
closer to those of the global fit. Depending on the nature
of the effect and its expected size, averaging can therefore be preferable for
null tests, although neither strategy undoes the masking, which requires
refitting the nuisance sources that absorb the effect, as we discuss next.
Simple post-processing is also computationally cheaper, so it is better to start
from it and add complexity as needed. For the remainder of the paper we will
focus on the simple post-processing strategy, and add comments on the averaged
strategy as needed.

\section{Computing the ratio and selecting the sources to refit}
\label{sec:practical}

To compute the masking and null ratios discussed in the last section, we need the two matrices $\Fc{b}{v}$ and
$\Fc{b}{cv}$. The first only involves the target source of interest and can be
computed and inverted directly, but the second requires marginalising over
\emph{all} sources in the data. For LISA, this means computing and
inverting a Fisher matrix over $\sim10^4$ resolvable sources, which is
computationally infeasible, especially given the large dynamic range that is typical of
GW Fisher matrices, which makes them ill-conditioned and unstable to invert numerically. As we now show, this
direct computation is not needed as the total masking can be estimated
from the posterior samples of the vacuum global fit.

\subsection{Estimating the total masking from the vacuum global fit posterior}
\label{sec:estimator}

The vacuum global fit already explores the joint posterior, and at
linear-signal order that posterior corresponds to the inverse Fisher matrix.
Recall the chain rule for the conditional matrices in Eq.~\eqref{eq:cGamma}. As
mentioned above, the first term can be computed directly. To obtain the second, we
need to isolate the beyond-vacuum direction from the target's own vacuum
absorption
\begin{equation}
\tilde u\;=\;\Pjperp{\Vs}\,\partial_{\thb}m \, ,
\qquad
\nip{\tilde u}{\tilde u}=\Fc{b}{v} \, ,
\label{eq:utilde}
\end{equation}
evaluated at $\thb=\bm 0$ and the best-fit vacuum parameters (we come back to
this choice of evaluation point in Sec.~\ref{sec:selection}). Then, for each
posterior sample $\bm\theta_s\sim p(\thc,\thv\mid d,M_{\rm vac})$ of the vacuum
global fit, we compute
\begin{equation}
\bm s_{\perp}(\bm\theta_s)=\nip{\tilde u}{m(\bm\theta_s)} \, .
\label{eq:sperpdef}
\end{equation}
Linearising around the posterior mean,
$m(\bm\theta_s)\simeq m(\bar{\bm\theta})+\partial_{\thc}m\,\delta\thc
+\partial_{\thv}m\,\delta\thv$, the vacuum contribution cancels because
$\nip{\tilde u}{\partial_{\thv}m}=\bm 0$ by construction, leaving
\begin{equation}
\bm s_{\perp}\simeq\mathrm{const}+\Gamma_{\mathrm{bc}\mid\mathrm{v}}\,\delta\thc ,
\qquad
\Gamma_{\mathrm{bc}\mid\mathrm{v}}=\nip{\tilde u}{\partial_{\thc}m}
=\F{bc}-\F{bv}\F{vv}^{-1}\F{vc} \, .
\label{eq:sperplin}
\end{equation}
The covariance of $\thc$ marginalised over $\thv$ is $\Fc{c}{v}^{-1}$, so again using
$\mathrm{Cov}[Ax]=A\,\mathrm{Cov}[x]\,A^{T}$
\begin{equation}
\mathrm{Cov}[\bm s_{\perp}]
=\Gamma_{\mathrm{bc}\mid\mathrm{v}}\,\Fc{c}{v}^{-1}\,
 \Gamma_{\mathrm{cb}\mid\mathrm{v}} \,  .
\label{eq:covsperp}
\end{equation}
which is exactly the term subtracted in the conditional
matrix~\eqref{eq:cGamma}. The fully marginalised matrix can then be obtained
from
\begin{equation}
\Fc{b}{cv}=\Fc{b}{v}
-\mathrm{Cov}_{\bm\theta_s\sim p(\thc,\thv\mid d,M_{\rm vac})}
 \bigl[\bm s_{\perp}\bigr]
\label{eq:sperpestimator}
\end{equation}
We stress that in this estimate no
Fisher matrix over the source catalogue is ever computed or inverted. The cost is
one waveform-derivative evaluation plus one inner product per posterior sample
evaluating the template $m(\bm\theta_s)$, which was already computed during the
first inference (and can in principle be saved). These are also the samples from which the
averaging strategy of Sec.~\ref{sec:marginalised} draws its nuisance
configurations, so the two reuse the same product of the first fit.

Since Eq.~\eqref{eq:sperpestimator} involves a sample variance, its Monte-Carlo
error can also be estimated. For $N_{\rm eff}$ effective posterior samples (the
number of samples divided by their integrated autocorrelation time), the relative error of
the covariance is $\sqrt{2/N_{\rm eff}}$, which propagates to
\begin{equation}
\frac{\delta\mathcal{R}}{\mathcal{R}}
\simeq\frac12\,
\frac{\mathrm{Cov}[\bm s_{\perp}]}{\Fc{b}{cv}}
\sqrt{\frac{2}{N_{\rm eff}}} \,
\label{eq:MCerror}
\end{equation}
for a single beyond-vacuum-GR parameter.

\subsection{Attributing the masking to a subset of nuisance sources}
\label{sec:subsets}

If both $\mathcal{R}$ and $\sqrt{\mathcal{R}_0}$ are very close to 1, no nuisance source can absorb the effect, and simple
post-processing is a safe strategy, with essentially no loss of detection
significance and bounds close to those of the global fit. How close to $1$ is close enough depends, ultimately, on
how much significance one is willing to give up in exchange for a cheaper refit.
What if this is not the case? The masking ratio is a single number for the whole
data set, so it quantifies how much detection significance post-processing
loses, but not to which nuisance sources it is lost, which is what we need in
order to choose the ones to refit jointly in a down-scaled reanalysis.

Writing $\thc_S$ for the parameters of a subset $S$ of nuisance sources, consider
their directions after the target source's vacuum directions are projected out,
\begin{equation}
\tilde V_S\;=\;
\bigl\{\Pjperp{\Vs}\,\partial_{\thc_j}\hc\bigr\}_{j\in S} \, .
\label{eq:Vtilde}
\end{equation}
The conditional matrix restricted to this subset is
\begin{equation}
\Gamma_{\mathrm{b}\mid\mathrm{v}S}=\Fc{b}{v}
-\Gamma_{\mathrm{b}\thc_S\mid\mathrm{v}}\,
 \Gamma_{\thc_S\thc_S\mid\mathrm{v}}^{-1}\,
 \Gamma_{\thc_S\mathrm{b}\mid\mathrm{v}}
=\Fc{b}{v}
-\nip{\tilde u}{\tilde V_S}\,
 \nip{\tilde V_S}{\tilde V_S}^{-1}\,
 \nip{\tilde V_S}{\tilde u} \, ,
\label{eq:GbvS}
\end{equation}
from which we compute the subset masking ratio
\begin{equation}
\mathcal{R}_S\;=\;
\sqrt{\frac{\Gamma_{\mathrm{b}\mid\mathrm{v}S}}{\Fc{b}{v}}}\, ,
\label{eq:RS}
\end{equation}
for a single beyond-vacuum parameter, with the obvious generalisation from
Eq.~\eqref{eq:Rdef} for $N_{\mathrm b}>1$. It has the same interpretation as
$\mathcal{R}$, as if the sources in $S$ were the only nuisance sources in the
data, and reduces to $\mathcal{R}$ itself when $S$ is the whole catalogue. We
denote the masking of a single source by
$\mathcal{R}_j\equiv\mathcal{R}_{\{j\}}$. Computing these ratios requires only
the waveform derivatives of the target and of the sources in $S$, so the matrix
to be inverted has the dimension of $S$ and can, in principle, be directly computed.

The residual masking left once $S$ has been included in the post-processing fit
is then
\begin{equation}
\mathcal{R}_{\mathrm{res}}\;=\;
\sqrt{\frac{\Fc{b}{cv}}{\Gamma_{\mathrm{b}\mid\mathrm{v}S}}}\;=\;
\frac{\mathcal{R}}{\mathcal{R}_S} \, .
\label{eq:factorise}
\end{equation}
So the masking factorises into the part due to $S$ and the part that remains
after $S$ is refitted, and since
$\Fc{b}{cv}\preceq\Gamma_{\mathrm{b}\mid\mathrm{v}S}\preceq\Fc{b}{v}$, both factors lie in $[0,1]$.

This residual ratio is the quantity that a down-scaled fit actually gives. Suppose we subtract from the data only the sources outside $S$, at the parameters found by
the vacuum global fit, and then fit the target together with the sources in $S$,
now with $\thb$ free. This is the two-step analysis of
Sec.~\ref{sec:StrategyPP}, except that the sources in $S$ are treated as the
global fit would treat them and only the rest are frozen and subtracted. Every
step of that derivation goes through with $\Vs$ replaced by
$\Vs\oplus \tilde V_S$, and therefore with $\Fc{b}{v}$ replaced by
$\Gamma_{\mathrm{b}\mid\mathrm{v}S}$. The analogues of
Eqs.~\eqref{eq:PPrecovered} and \eqref{eq:sig_loss} for a single beyond-vacuum parameter are
\begin{equation}
\bigl\langle(\thb_{\rm bf})^{S}\bigr\rangle
=\Gamma_{\mathrm{b}\mid\mathrm{v}S}^{-1}\,\Fc{b}{cv}\,\thb_{\mathrm{eff}}
=\mathcal{R}_{\mathrm{res}}^{2}\,\thb_{\mathrm{eff}} \, ,
\qquad
\sqrt{2S^{S}}=\mathcal{R}_{\mathrm{res}}\,\sqrt{2S^{\mathrm{GF}}} \, .
\label{eq:downscaled}
\end{equation}
The recovered value is shrunk by
$\mathcal{R}_{\mathrm{res}}^{2}$ and the quoted uncertainty
$\Gamma_{\mathrm{b}\mid\mathrm{v}S}^{-1/2}$ is too tight by
$\mathcal{R}_{\mathrm{res}}$, exactly as in Sec.~\ref{sec:comparison} but with
the residual ratio in place of the total one. The corresponding null statement
is that the down-scaled analysis has null ratio
\begin{equation}
\mathcal{R}_{0}^{\,S}
=\frac{\Tr\bigl(\Gamma_{\mathrm{b}\mid\mathrm{v}S}^{-1}\Fc{b}{cv}\bigr)}
      {N_{\mathrm b}} \, ,
\label{eq:R0S}
\end{equation}
so a bound set in the absence of any effect is too tight by the same mechanism,
and by the same factor when $N_{\mathrm b}=1$.

The two limits recover the strategies already studied. For $S=\varnothing$ we
have $\mathcal{R}_S=1$ and $\mathcal{R}_{\mathrm{res}}=\mathcal{R}$, and
Eq.~\eqref{eq:downscaled} is simple post-processing. Conversely, for $S$ equal to the whole
catalogue, $\mathcal{R}_S=\mathcal{R}$ and $\mathcal{R}_{\mathrm{res}}=1$, and
it is the global fit.

\subsection{Selecting the sources}
\label{sec:selection}

We are then looking for the smallest set of sources worth refitting together with
the target source of interest. Let $\tilde u_{\bar S}$ denote the part of
$\tilde u$ that the sources in $S$ cannot reproduce, so that
$\nip{\tilde u_{\bar S}}{\tilde u_{\bar S}}=\Gamma_{\mathrm{b}\mid\mathrm{v}S}$.
Starting from $S$ empty, at each step:
\begin{enumerate}[leftmargin=1.6em]
\item For each candidate $j$ not included in $S$, project out from $\tilde V_j$
the directions already covered by $S$,
$\tilde V_j^{\,\prime}=\Pjperp{\mathrm{span}\,\tilde V_S}\,\tilde V_j$, and
compute how much of $\tilde u_{\bar S}$ it accounts for,
\begin{equation}
\Delta_j=\nip{\tilde u_{\bar S}}{\tilde V_j^{\,\prime}}\,
         \nip{\tilde V_j^{\,\prime}}{\tilde V_j^{\,\prime}}^{-1}\,
         \nip{\tilde V_j^{\,\prime}}{\tilde u_{\bar S}} \, .
\label{eq:Deltaj}
\end{equation}
\item Add the candidate with the largest $\Delta_j$ to $S$ and update
$\tilde u_{\bar S}$ by removing this contribution.
\item Evaluate
$\mathcal{R}_S=\sqrt{\nip{\tilde u_{\bar S}}{\tilde u_{\bar S}}/\nip{\tilde u}{\tilde u}}$
and $\mathcal{R}_{\mathrm{res}}=\mathcal{R}/\mathcal{R}_S$.
\item Stop when the single measure of Eq.~\eqref{eq:Rmin}, computed for the
residual masking, $\min(\mathcal{R}_{\rm res},\sqrt{\mathcal{R}^S_0})$ with
Eqs.~\eqref{eq:factorise} and~\eqref{eq:R0S}, is close enough to 1. For
$N_{\rm b}=1$ it is simply $\mathcal{R}_{\rm res}$.

\end{enumerate}
This is a greedy algorithm since at each step it adds the source that recovers the
most masking, without revisiting earlier choices. For $N_{\rm b}>1$, $\Delta_j$
is an $N_{\rm b}\times N_{\rm b}$ matrix, and the subset ratios generalise to
the eigenvalues of $\Gamma_{{\rm b}|{\rm v}S}^{-1}\Fc{b}{cv}$, as in
Eq.~\eqref{eq:chi2pp}. When a specific beyond-vacuum model is tested, both are
projected along the direction it predicts, which reduces the selection to the
single-parameter case. For a generic reanalysis, we instead add at each step
the candidate that most increases the smallest of these eigenvalues, and stop
when its square root is close enough to one, so that the selected set is
adequate for an effect in any direction. Once $S$ is selected, we subtract from the data only the sources outside it and
refit the target source jointly with the sources in $S$, now with $\thb$ free.
Since $\tilde V_j$ is projected perpendicular to the sources already chosen,
$\Delta_j$ measures what candidate $j$ adds to the selected set. The projection can be updated sequentially as each
source is added to $S$, since only its own direction has to be projected out of each
remaining candidate and of $\tilde u_{\bar S}$. Every step therefore involves
only inverses of matrices the size of a single source's parameter space not yet included in S, as in
Eq.~\eqref{eq:Deltaj}, however large $S$ has become. The same selection can be
used in the averaging strategy of Sec.~\ref{sec:marginalised}. The $\Delta_j$
depend only on the target and nuisance directions, not on how the sources
outside $S$ are subtracted, and for a single beyond-vacuum parameter
$\bar{\mathcal{R}}$ is a monotonic function of $\mathcal{R}$, so the ranking is the same as in simple
post-processing and only the stopping threshold changes.

In the first iteration of this algorithm, with $S$ empty, we obtain the
individual ratios $\mathcal{R}_j$, with which we can rank the catalogue from the
most to the least problematic source. Some extra care is needed when nuisance
sources are strongly correlated among themselves. If two sources can reproduce
the same part of $\tilde u$, keeping one of them is enough. However, it can also
happen that $\tilde u$ is only well reproduced by a combination of the two, and
then each source is relatively harmless individually ($\mathcal{R}_j\simeq1$) even though the
pair is not. This can only happen for sources that are correlated with each other. If their directions are orthogonal, $(\tilde V_j|\tilde V_k)=0$, their contributions simply add, $\Delta_{\{j,k\}}=\Delta_j+\Delta_k$, whereas two nearly degenerate
sources can together reproduce a direction that neither reproduces alone. To account for this, we can also include small groups of mutually correlated
sources as single candidates in step 1, which can be identified from the correlations in the posterior of the first vacuum global fit. We will see an example of this in the next section where we apply our method to a simple toy model.

Finally, recall that all derivatives are evaluated at $\thb=\bm 0$, since that is what the vacuum global fit provides. If the beyond-vacuum GR effect is very strong, it can enter the waveform non-linearly, and the ratios and the ranking of
nuisance sources will then depend on the true value of $\thb_0$. This is not in
conflict with the linear-signal approximation, but means that a single tangent
plane may not extrapolate well across the full range of $\thb$ allowed. Part
of this is already captured at $\thb=\bm 0$, because the vacuum global fit
absorbs part of the effect into the vacuum parameters of the target, so its best
fit follows the deviated signal. The null ratio is
not affected by this, since $\thb=\bm 0$ is the correct expansion point when no
effect is present. For the masking ratio and the selection of $S$, one can
instead repeat the procedure above at a few values of $\thb$, take for $S$ the
union of the sources flagged anywhere in the sweep, and report the smallest
$\mathcal{R}$ obtained. Alternatively, the expansion point can be updated iteratively by evaluating the
ranking at the current estimate of $\thb$, refit with the selected sources,
and repeat until the selected set no longer changes. In both cases the vacuum
parameters must move with $\thb$, that is, be taken from a fit at that value of
$\thb$. Evaluating the derivatives at fixed vacuum parameters while varying
$\thb$ moves the template away from the data, and the ratios then vary
erratically. We will see this in play in our toy model example.

\section{A toy model}
\label{sec:ToyModel}

\begin{figure}[t]
\centering
\includegraphics[width=1.02\textwidth]{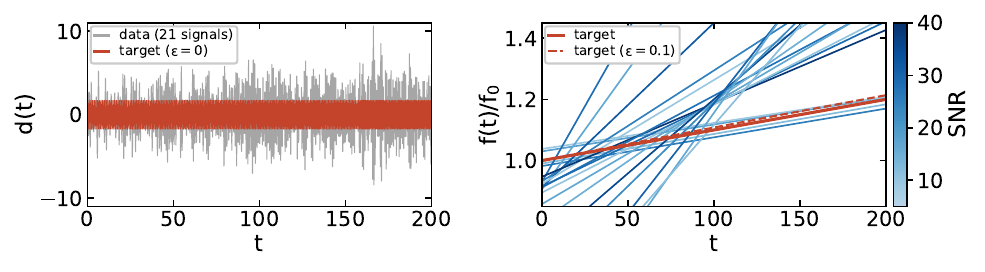}
\caption{The toy model dataset. \emph{Left:} the (noiseless) time-domain data stream,
with the target source, with $\rm SNR =50$, highlighted. \emph{Right:} frequency
tracks of the $21$ chirping signals, coloured by SNR. The target is flanked by
two pairs of nearly parallel tracks, group A below and group B above, while the
remaining $16$ sources cross its track during the observation. The dashed line
is the target's track when the beyond-vacuum effect of Eq.~\eqref{eq:toyeps} is
present with $\varepsilon=0.1$.}
\label{fig:toytracks}
\end{figure}

We now apply the procedure of the previous section to a toy model, and validate
it against full Markov-chain Monte Carlo (MCMC) inference. The dataset is
illustrated in Fig.~\ref{fig:toytracks}. It consists of a noiseless data stream of $21$ linearly chirping sinusoids
\begin{equation}
d(t)=\sum_{j=0}^{20} A_j
\cos \bigl(\phi_j+2\pi f_j\,t+\pi \dot f_j\,t^{2}\bigr) \, ,
\label{eq:toydata}
\end{equation}
where $A_j$ and $\phi_j$ are the amplitude and initial phase of source $j$, and
$f_j$ and $\dot f_j$ are constant for the $20$ nuisance sources. All
quantities are in dimensionless units with $f_0=1$ and $\dot f_0=10^{-3}$, and
the noise has a flat power spectral density, $S(f)=S_0$. The data are sampled
for $-100\le t\le200$, and the noise-weighted inner product carries a window
$w(t)$ equal to one for $0\le t\le100$ and falling to zero with cosine tapers
over $100$ time units on either side, so that the effective observation time is
$T=\int w\,\mathrm{d}t=200$. Source $j=0$ is the target source of
interest, with $\rm SNR=50$, and its frequency evolution can carry a non-linear
chirp term
\begin{equation}
\frac{\mathrm{d}f}{\mathrm{d}t}
=\dot f_0\left[1+\varepsilon\left(\frac{f}{f_0}\right)^{\!n}\right] ,
\label{eq:toyeps}
\end{equation}
in which case its phase is obtained by integrating Eq.~\eqref{eq:toyeps}. This
plays the role of the beyond-vacuum GR effect in this toy model, with
$\varepsilon=0$ recovering vacuum. To keep the illustration simple we fix
$n=-4$, so that the beyond-vacuum subspace has dimension one. We experimented
with other choices and the qualitative picture below does not change. As
discussed in Sec.~\ref{sec:comparison}, this is similar in spirit to the
parametrised tests of GR used in current GW analyses, where the order of the
deviation is fixed and only its amplitude is fitted.

The remaining $20$ sources form the nuisance population. Four of them lie close
to the target and chirp almost in parallel with it. We divide them into two groups. Sources $\{1,2\}$, which we
label group A, at frequency offsets $\Delta f=-0.010$ and $-0.018$ with
$\dot f=0.97 \,\dot f_0$ and $0.94\,\dot f_0$; and sources $\{3,4\}$, group B, at
$\Delta f=+0.030$ and $+0.038$ with $\dot f=0.88 \,\dot f_0$ and $0.85\,\dot f_0$. The
other $16$ chirp much faster and cross the target's track at some point during
the observation. Their SNRs range from $9$ to $38$.

In all MCMC inference, we sample the posterior with the \texttt{Eryn}
sampler~\cite{Karnesis:2023ras}, using 128 walkers initialised in a Gaussian of width $0.05\,\sigma$ around the
maximum-likelihood point of each analysis.  We discard at least $5\times10^3$ iterations as burn-in and check convergence
with the integrated autocorrelation time, which for the slowest direction of the
84- and 85-dimensional fits is a few hundred stored steps, while the chains have
$10^4$--$5\times10^4$ stored steps.
Here, $\sigma$ is the marginal Fisher uncertainty of each parameter, obtained by inverting the full joint
Fisher matrix of the whole population, which in this case can be computed
directly given the simplicity of the source catalogue. The priors are uniform
and span $\pm8\,\sigma$ around the same maximum-likelihood point, wide enough that
they do not truncate the posterior of $\varepsilon$.

\subsection{Vacuum Injection}
\label{sec:ToyNull}

\begin{figure}[t]
\centering
\includegraphics[width=1\textwidth]{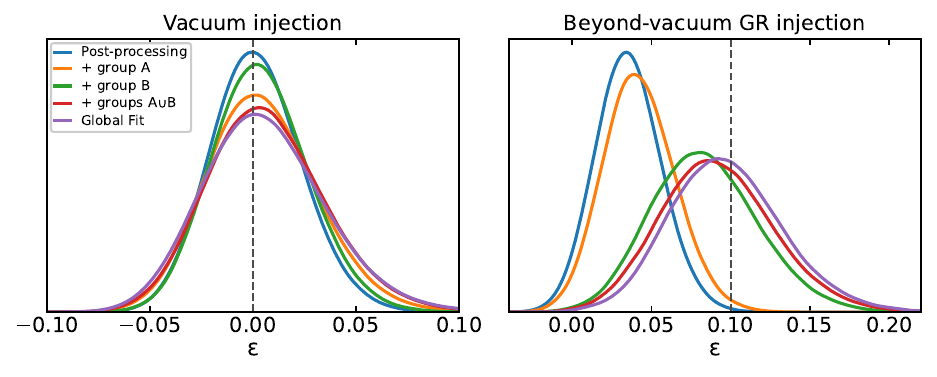}
\caption{Marginal MCMC posteriors of the beyond-vacuum parameter $\varepsilon$
under the different strategies: simple post-processing, in which all nuisance
sources are subtracted and only the target is refitted; the down-scaled fits in
which group A, group B, or both are refitted jointly with the target source; and the
full global fit. \emph{Left:} vacuum dataset, $\varepsilon_{0}=0$. All posteriors
are centred on the truth (vertical dashed line), and only their widths differ.
Post-processing is the narrowest, so the upper bound it implies is tighter than
the data support, and freeing each group widens the posterior by the
$1/\mathcal{R}_S$ of Table~\ref{tab:nullrank}, until with both groups freed it is
almost indistinguishable from the global fit. \emph{Right:} with a beyond-vacuum
effect present, $\varepsilon_{0}=0.1$. Post-processing is now shifted towards
vacuum as well, by the factor $\mathcal{R}^{2}$ of Eq.~\eqref{eq:PPmap}, so that
it recovers a weaker effect at lower significance, in this case compatible with
the null hypothesis. Freeing the groups moves the posterior back towards the
truth, but note that here it is group B that recovers most of the effect, whereas
in the vacuum case it was group A that carried most of the masking.}
\label{fig:toynull}
\end{figure}

We first consider the case where no deviation is present, so that
$\varepsilon_{0}=0$. The question is then what upper bound each strategy gives,
and whether the null ratio predicts the difference between them. Since no effect
is in the data, $\thb=\bm 0$ is the correct expansion point and the caveats
discussed at the end of Sec.~\ref{sec:selection} do not apply. From the direct
Fisher matrix, the total masking evaluates to $\mathcal{R}=0.78$, which for $N_{\rm b} = 1$ is also $\sqrt{\mathcal{R}_0}$. The posterior
estimator of Eq.~\eqref{eq:sperpestimator}, using $\sim7\times10^{3}$ effective
samples of an $84$-dimensional vacuum global fit of the same data is in good agreement, giving
$\mathcal{R}=0.784\pm0.004$, where the uncertainty is the Monte-Carlo error of
Eq.~\eqref{eq:MCerror}.

We then ran the two strategies. The global fit, with $\varepsilon$ free together
with all $20$ nuisance sources for a total of $85$ parameters, gives
$\sigma_{\varepsilon}^{\rm GF}=0.030$, to be compared with the Fisher prediction
$\Fc{b}{cv}^{-1/2}=0.028$. Target-only post-processing,
that is a vacuum global fit
followed by a refit of the target alone with $\varepsilon$ free, gives
$\sigma_{\varepsilon}^{\rm PP}=0.022$, matching $\Fc{b}{v}^{-1/2}=0.022$. The
corresponding posteriors are shown in the left panel of Fig.~\ref{fig:toynull},
and the full 2-D marginals for the target source are given in
Appendix~\ref{app:JointPosteriorNull}. Both are unbiased, as they must be when no
extra effect is present in the data and there is no noise, but the ratio of the
reported uncertainties is $0.74$, in good agreement with the predicted
$\mathcal{R}=0.78$.

We then ranked the nuisance sources with the algorithm of
Sec.~\ref{sec:selection}, and the results are collected in
Table~\ref{tab:nullrank}. All $16$ crossing sources are individually harmless,
with $\mathcal{R}_j\ge0.997$ regardless of their SNR, although collectively they
still reach $\mathcal{R}_S=0.97$. The four nearly parallel sources are more
problematic,  with
$\mathcal{R}_A=0.85$ (group A) and $\mathcal{R}_B=0.96$ (group B). Group B also illustrates the joint masking discussed in
Sec.~\ref{sec:selection}. Its two sources are strongly correlated with each
other, and although each is individually almost harmless
($\mathcal{R}_j\simeq0.99$), together they absorb almost twice as much as the
two separately. These subset ratios predict how much of the masking each down-scaled fit
recovers. Freeing group A together with the target in the refit should raise the residual ratio
from $\mathcal{R}=0.78$ to $\mathcal{R}_{\rm res}=0.92$, while freeing group B to
$0.82$, and freeing both to $0.96$. We checked this by repeating the second step
of post-processing in each case, shown also in the left panel of
Fig.~\ref{fig:toynull} (and in Appendix~\ref{app:JointPosteriorNull} for the marginal 2D posteriors). The
resulting $95\%$ bounds are $\varepsilon<0.050$ with group A freed,
$\varepsilon<0.043$ with group B, and $\varepsilon<0.056$ with both, against
$\varepsilon<0.039$ for target-only post-processing and $\varepsilon<0.057$ for the
global fit. The bounds therefore go towards the global-fit value in the order
the ratios predict, with the posterior widths following the expected
$1/\mathcal{R}_S$ inflation (to within $5\%$), and the refit with both groups freed
is almost indistinguishable from the global fit. We highlight that in this case it
is group A that carries most of the masking compared to group B. As we show next, this ordering is
inverted once a strong beyond-vacuum GR effect is actually present.

\begin{table}[t]
\centering
\begin{tabular}{lccccc}
\toprule
 & $\Delta f$ & $\dot f/\dot f_0$ & SNR & $\mathcal{R}_j$ & $\mathcal{R}_S$ \\
\midrule
group A                 &          &              &           &             & $0.85$ \\
\quad source 1          & $-0.010$ & $0.97$       & $12$      & $0.87$      &        \\
\quad source 2          & $-0.018$ & $0.94$       & $25$      & $0.98$      &        \\
\addlinespace
group B                 &          &              &           &             & $0.96$ \\
\quad source 3          & $+0.030$ & $0.88$       & $18$      & $0.99$      &        \\
\quad source 4          & $+0.038$ & $0.85$       & $9$       & $0.99$      &        \\
\addlinespace
crossers ($16$ sources) & ---      & $2.4$--$9.9$ & $9$--$38$ & $\ge 0.997$ & $0.97$ \\
\midrule
groups A$\,\cup\,$B     &          &              &           &             & $0.81$ \\
all $20$ sources $(=\mathcal{R})$ & &            &           &             & $0.78$ \\
\bottomrule
\end{tabular}
\caption{Ranking of the nuisance catalogue for the vacuum injection, at the null point ($\thb=\bm 0$).
$\Delta f$ is the frequency offset from the target at $t=0$ and
$\dot f/\dot f_0$ the chirp-rate ratio, $\mathcal{R}_j$ is the individual masking
ratio of each source, and $\mathcal{R}_S$ the collective ratio of the group it
belongs to. The four nearly parallel sources dominate, with no correlation with
SNR, while the $16$ crossers are individually harmless
($\mathcal{R}_j\ge0.997$) but reach $\mathcal{R}_S=0.97$ together. The last row
is the full catalogue, recovering the total $\mathcal{R}$.}
\label{tab:nullrank}
\end{table}

\subsection{Beyond-vacuum GR injection}
\label{sec:ToyBV}

We now inject $\varepsilon_0=0.1$, which corresponds to $2.8\,$rad of
dephasing relative to vacuum . The global
fit recovers $\varepsilon=0.097\pm0.035$, containing the truth and excluding
vacuum at $2.8\sigma$, while post-processing recovers
$\varepsilon=0.036\pm0.021$. It is biased towards vacuum, excluding the true value at
$3\sigma$, and inconsistent with vacuum at only $1.7\sigma$. This is the
loss in reported significance of Eq.~\eqref{eq:sig_loss}, which predicts
$\mathcal{R}\sqrt{\lambda^{\rm GF}}=1.7\sigma$. Evaluated at $\varepsilon=0.1$,
the masking ratio is $\mathcal{R}=0.66$, so post-processing should recover
$\mathcal{R}^{2}\,\varepsilon_0=0.043$, compatible with the measured
$0.036\pm0.021$. Freeing the different subsets of nuisance sources in the second
inference step brings the recovered value back towards the truth according to
the residual ratio of Eq.~\eqref{eq:factorise}. Table~\ref{tab:ladder} compares
the predictions $\mathcal{R}_{\mathrm{res}}^{2}\,\varepsilon_0$ with the MCMC
results for the different strategies, and the right panel of
Fig.~\ref{fig:toynull} shows the corresponding posteriors for $\varepsilon$.
Every step agrees with the prediction to within a small fraction of the posterior width. Note that
freeing only group B already almost recovers the global-fit result, while
freeing group A alone changes little, the opposite ofthe ranking of the vacuum injection in
Table~\ref{tab:nullrank}.

The bias in the post-processing strategies is not limited to $\varepsilon$.
Figure~\ref{fig:cornerbv} in Appendix~\ref{app:JointPosteriorBGR} shows the full
2D-posteriors of the target source for each strategy. In target-only post-processing,
the target's own chirp rate is displaced as well, by
$\Delta\dot f_0\simeq6\times10^{-5}$, a bias more than $3\times$  its own posterior
width, along the $\varepsilon$--$\dot f_0$ degeneracy visible in the
corresponding panel. The initial frequency follows along the same degeneracy, by
about two of its own widths, while the amplitude $A$ moves by less than
one. This is expected, as the part of the effect absorbed by the nuisance
sources is subtracted from the residual, and the second fit compensates for what
is missing by moving whichever vacuum parameters of the target can mimic it.
Freeing group B removes almost all of these displacements at once, whereas
freeing group A leaves $\varepsilon$ and $\dot f_0$ essentially where simple
post-processing left them.

The nuisance sources are affected in the same way, as shown in
Figs.~\ref{fig:cornerAbv} and~\ref{fig:cornerBbv} of
Appendix~\ref{app:JointPosteriorBGR}, which also show the first step of
post-processing. In target-only post-processing
their parameters are never refitted. They stay frozen at the values returned by
the vacuum global fit, which absorbed part of the beyond-vacuum GR effect and are biased by
Eq.~\eqref{eq:Deltac_dh}. In units of the width of the first-step
posterior, the best-fit chirp rates of the two sources of
group B are displaced by $2.4\sigma$ and $1.1\sigma$ respectively, the
parameters of group A by at most $0.5\sigma$, and those of the $16$ crossing
sources by at most $0.3\sigma$, with a mean of $0.1\sigma$. Refitting
group B together with the target brings its parameters back to within
$0.3\sigma$ of the injection. The sources that do
the masking are therefore the same ones whose parameters are most biased by
it, and the ranking of Sec.~\ref{sec:selection} is able to identify them in this toy model.

This example also illustrates the role of the expansion point. An analyst of
these data does not have the ranking of Table~\ref{tab:nullrank}, which belongs
to the vacuum injection. The ratios are instead evaluated at $\thb=\bm 0$ and the
best fit of the vacuum global fit of these data, as prescribed in
Sec.~\ref{sec:estimator}. That fit has absorbed part of the effect into the
target's own vacuum parameters, so its template already follows the deviated
track, and there $\mathcal{R}=0.66$, $\mathcal{R}_A=0.93$ and
$\mathcal{R}_B=0.72$. These are essentially the values at $\varepsilon=0.1$ in
Table~\ref{tab:ladder}, the opposite of the vacuum-injection ranking, and the
MCMC results agree with them. Non-linearity is also visible in the posteriors,
which are not completely Gaussian. Nonetheless,
the linear-signal predictions of Table~\ref{tab:ladder} recover the measured
means to within a fraction of the posterior width, so the approximation is
adequate for ranking and selecting sources even where it is no longer exact for
the shape of the posterior.

Applied at this point, the selection of Sec.~\ref{sec:selection} adds group B
first ($\mathcal{R}_{\rm res}=0.91$) and then group A
($\mathcal{R}_{\rm res}=0.96$). With $\mathcal{R}_{\rm res}\ge0.9$ as the
threshold, refitting group B alone recovers $\varepsilon=0.084\pm0.034$, within
$0.4\sigma$ of the global fit. With $\mathcal{R}_{\rm res}\ge0.95$, refitting both
groups recovers $\varepsilon=0.092\pm0.036$, within $0.15\sigma$, with $16$
nuisance parameters refitted instead of $80$. Re-evaluating the ranking at the
values recovered by either refit leaves the selection unchanged, so the
iterative update of Sec.~\ref{sec:selection} stops after the first refit. Had
the ranking been taken from the vacuum injection instead, it would have pointed
to group A, whose refit recovers only $\varepsilon=0.043\pm0.023$. The expansion
point must therefore follow the data, and when it is swept or updated the vacuum
parameters have to move with $\thb$. At fixed vacuum parameters, $\mathcal{R}$
changes from $0.61$ to $0.80$ between $\varepsilon=0.09$ and $0.11$, because the
target's template then drifts out of phase with the data, and with it its
overlap with the nearly parallel nuisance sources.

\begin{table}[t]
\centering
\begin{tabular}{lccccc}
\hline
subset $S$ & $\mathcal{R}_S(0)$ & $\mathcal{R}_S(0.1)$ & $\mathcal{R}_{\rm res}$ &
predicted $\varepsilon$ & measured $\varepsilon$ \\
\hline
post-processing ($S=\varnothing$) & $1.00$ & $1.00$ & $0.66$ & $0.043$ & $0.036\pm0.021$ \\
$+$ group A $\{1,2\}$             & $0.93$ & $0.94$ & $0.70$ & $0.049$ & $0.043\pm0.023$ \\
$+$ group B $\{3,4\}$             & $0.72$ & $0.71$ & $0.93$ & $0.087$ & $0.084\pm0.034$ \\
$+$ groups A$\,\cup\,$B           & $0.69$ & $0.69$ & $0.96$ & $0.091$ & $0.092\pm0.036$ \\
global fit ($S=$ all $20$)        & $0.66$ & $0.66$ & $1.00$ & $0.100$ & $0.097\pm0.035$ \\
\hline
\end{tabular}
\caption{Down-scaling ranking for the toy model at $\varepsilon_0=0.1$. Each row
corresponds to freeing the subset $S$ of Table~\ref{tab:nullrank} together with
the target source in the second refit step of post-processing, from simple
post-processing ($S=\varnothing$) to the full global fit. The subset ratio is
quoted at two expansion points: $\mathcal{R}_S(0)$ at $\thb=\bm 0$ and the
vacuum best fit of these data, and $\mathcal{R}_S(0.1)$ at the injected
parameters, which is used here. The residual
masking follows from Eq.~\eqref{eq:factorise} and the predicted $\varepsilon$
from Eq.~\eqref{eq:downscaled}. The measured values are posterior means and
standard deviations of the corresponding MCMC analyses. Both give the same
ranking, the opposite of the vacuum injection in Table~\ref{tab:nullrank}.}
\label{tab:ladder}
\end{table}

\subsection{Averaged post-processing}
\label{sec:ToyMarg}

\begin{figure}[t!]
\centering
\includegraphics[width=0.95\textwidth]{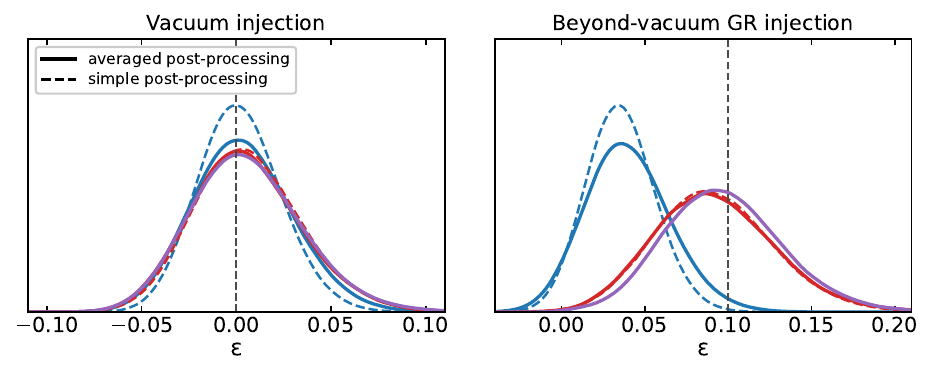}
\caption{The averaged post-processing strategy of Sec.~\ref{sec:marginalised}
in the toy model, for the target-only refit ($S=\varnothing$) and the
down-scaled fit with groups A and B freed, against the global fit, with the
same colour scheme as in Fig.~\ref{fig:toynull}. Solid curves are the averaged
posteriors, obtained by drawing $128$ nuisance configurations per analysis
from the first-step vacuum posterior, rerunning the second step for each
draw, and concatenating the chains. Dashed curves are the corresponding simple
(best-fit-subtraction) posteriors of Fig.~\ref{fig:toynull}. \emph{Left:} vacuum injection. Averaging widens each posterior towards the
global-fit width by $\sqrt{2-\mathcal{R}_{\rm res}^{2}}$,
Eq.~\eqref{eq:margscalar} with $\mathcal{R}\to\mathcal{R}_{\rm res}$, so the
reported bound becomes better calibrated, and the more sources are freed the
smaller the correction.
\emph{Right:} beyond-vacuum injection. The means are essentially unchanged
(Sec.~\ref{sec:marginalised}), so averaging does not undo the masking of
Eq.~\eqref{eq:PPmap}. The recovered value stays at
$\mathcal{R}_{\rm res}^{2}\,\varepsilon_0$ while the detection significance drops by
$\sqrt{2-\mathcal{R}_{\rm res}^{2}}$, Eq.~\eqref{eq:margscalar}. }
\label{fig:margtoy}
\end{figure}

Finally, we validate the averaged post-processing strategy of
Sec.~\ref{sec:marginalised} on the same two datasets, for the target-only
refit and for the down-scaled fit with groups A and B freed. In each case, we
draw $128$ nuisance configurations from the first-step vacuum posterior
chains, subtract each, rerun the second inference step, and concatenate the
resulting chains, for $512$ per-draw analyses in total. Each per-draw refit is a shorter
\texttt{Eryn} run, with $64$ walkers and $1200$ stored iterations after $500$ of
burn-in, started at that draw's maximum-likelihood point. A control run five
times longer, for one draw of the vacuum injection, changes the per-draw width by
at most $4\%$ and its mean by less than $0.1\sigma$. The averaged
posteriors are shown in Fig.~\ref{fig:margtoy}, and the ratios between their
widths and the simple post-processing ones follow Eq.~\eqref{eq:margscalar} to $\lesssim 2\%$ for
both the vacuum and the beyond-vacuum injections. For the target-only refit,
averaging widens the posterior from $\sigma_\varepsilon=0.022$ to $0.027$ in
the vacuum injection (predicted $0.026$) and from $0.021$ to $0.025$ in the
beyond-vacuum one (predicted $0.026$), while with both groups freed the
averaged width is within $\lesssim 4\%$ of the global-fit one in both cases. The two
consequences anticipated in Sec.~\ref{sec:comparison} are visible directly.
In the vacuum injection, the bound becomes better calibrated, and the more
nuisance sources are freed the less there is left to correct. The width
difference with respect to the global fit falls from $26\%$ to $11\%$ for the
target-only refit and from $4\%$ to $2\%$ with groups A and B freed. In the
beyond-vacuum injection the significance of the detection degrades, with the recovered deviation
dropping from $1.7\sigma$ to $1.6\sigma$, slightly above the predicted
$1.4\sigma$ because the averaged mean sits $0.2\sigma$ above the
best-fit-subtraction one, which we attribute to a small departure from
Gaussianity of the first-step posterior.

\section{Discussion}
\label{sec:Disc}

We have presented a criterion to decide which overlapping signals must be
reanalysed jointly when one wants to test a different model for some target source than the
one used in an initial global fit so that any extra effect in the new model is not absorbed and masked by other nuisance signals. This criterion is obtained by computing the overlap between the directions that a group of nuisance sources spans in data space and the
direction of an extra effect absent from the first model.
The absorption is captured by two numbers. The masking ratio $\mathcal{R}$
of Sec.~\ref{sec:comparison} is the factor multiplying the
reported detection significance when an effect is present, so that a deviation the global fit would
establish at $n_\sigma$ is reported by post-processing at
$\mathcal{R}\,n_\sigma$. The null ratio $\mathcal{R}_0$ quantifies the
miscalibration when no effect is present. It can be interpreted as the factor by which the
posterior variance shrinks,
so the bounds that post-processing reports are tighter
than those from the full global fit by $\sqrt{\mathcal{R}_0}$. For an effect of unknown size, the loss lies between the two, so
the smaller of
$\mathcal{R}$ and $\sqrt{\mathcal{R}_0}$ serves as a single,
conservative
measure. For a single beyond-vacuum parameter,
the two
coincide.
Both can be computed by reusing the posterior samples of the initial vacuum global fit. We also
construct an algorithm that ranks the nuisance sources by how strongly they mask a
specific beyond-vacuum effect in a given target source. The rationale behind this is to keep the
smallest subset that reproduces the inference of the full global fit, and to
refit only those sources jointly with the target in a down-scaled reanalysis.  The linear-signal predictions for the uncertainties, the ratios,
the recovered values and the down-scaled bounds were confirmed, to within a
fraction of the posterior width, by full MCMC
analyses on a toy model that included one target source with a possible deviation and 20 nuisance sources.

Our methods are directly applicable to LISA data-analysis pipelines. The LISA
Science Ground Segment currently plans to release catalogues of sources
characterised with a global fit using vacuum GR  only. The
method described here could be used to assess for which sources and which
types of beyond-vacuum GR deviations robust constraints can be obtained by
post-processing only the target source, or the target source together with a
small number of nuisance sources. We only need the overlaps of
Eq.~\eqref{eq:sperpdef} between the beyond-vacuum direction of the target and
the templates at the posterior samples of the vacuum global fit. For a set of standard
beyond-vacuum models (e.g. scalar-tensor theories, environmental effects), these could be computed during the first inference and
released alongside the catalogue, and for any other model they can be computed
from the released posterior samples, so that any user can compute the masking
of their own beyond-vacuum model without rerunning the global fit. It will be
possible to test this using forthcoming mock datasets. The LISA Distributed Data
Processing Centre (DDPC) has generated a \textit{Mojito} dataset, which contains
the major LISA source classes together with non-stationary noise. This dataset
will be analysed with the global-fit pipelines currently under development by
the DDPC and the NASA Ground Segment, and the resulting posteriors will be
released. The \textit{Mojito}
dataset contains no beyond-vacuum GR physics, so an analysis of such effects
falls precisely into our null case, and the quantities to compute are
$\mathcal{R}_0$ and the per-source rankings. Once the vacuum pipelines are
robust and their results available, our estimator can be applied to assess
whether post-processing is a viable strategy for the scientific interpretation
of the public data products, for which sources a down-scaled reanalysis is
required instead, and under what circumstances the only option is to redo the
entire global fit. Through this experience, and the analysis of future common
datasets, we will learn how to obtain robust constraints on beyond-vacuum GR
effects in a computationally efficient way from the real LISA data.

Although our toy model did not include systematic errors, the formalism accounts
for them. Waveform model error enters both strategies through the same effective
parameter $\thb_{\mathrm{eff}}$, displacing what both tests measure and driving
the false-positive term of Eq.~\eqref{eq:maskgf_u}. The condition for a
threshold to remain meaningful, $\lambda_{\delta m}=(\bm Z_{\delta m}^{\mathrm{GF}})^{T}\Fc{b}{cv}^{-1}\bm
Z_{\delta m}^{\mathrm{GF}}\ll 1$, is then a requirement not on the overall
accuracy of the waveform model but on its projection onto the beyond-vacuum
direction. This is an important topic to explore further, but will be sensitive to the assumed
waveform error model, and in particular its projection onto the beyond-vacuum-GR space, and so it
must be done carefully.

One of the main limitations of our method is the use of the linear-signal
approximation~\cite{Vallisneri:2007ev}, which enters both in approximating the
vacuum manifold by its tangent plane and in characterising the first fit by a
point estimate at the maximum of the likelihood. Curvature, priors, and
subtraction at a posterior point estimate rather than at a maximum all degrade
the exact relation between the two strategies, and we already saw in
Sec.~\ref{sec:ToyBV} that the ranking itself depends on the expansion point
once the effect is large enough. Evaluating the ranking at the vacuum best fit
of the data already accounts for part of this, because the vacuum fit absorbs
part of the effect into the target's own parameters, and in the toy model this
alone identifies the right sources. Sweeping the expansion point, or updating it
at the best fit of each refit, with the vacuum parameters moving with it, can
account for the rest. It is also important to note that the final reanalysis itself
does not rely on the linear-signal approximation, which only enters the
selection of the sources. If the approximation failed to identify an important
nuisance source, this could be compensated by a more stringent threshold. In
addition, estimating the masking ratio from the posterior samples themselves,
as described in Sec.~\ref{sec:estimator}, should make it robust to some
deviations from the linear-signal approximation. This is one of the points that
can be tested using the DDPC common datasets.

Another aspect to explore more carefully with the common datasets is when and how
to average in post-processing, i.e. to average over samples from the first fit
instead of subtracting its best fit. We discussed this strategy in
Secs.~\ref{sec:marginalised} and~\ref{sec:comparison}, where we showed that it
reduces the significance of a detection by at most a factor of $\sqrt2$~\eqref{eq:chi2pool}, while bringing the upper limits of non-detections
closer to those of the global fit~\eqref{eq:R0ULbar}. This comes at the
price of increased computational cost (one refit per posterior draw). However,
it may have the advantage of increased robustness against the modelling
approximations we were just discussing, and of better-calibrated uncertainties
on the vacuum parameters, whose means it leaves unchanged at linear order. Using
the common dataset analyses, it will be possible to compare the simple and
averaged strategies in a realistic setting, and hence identify the right
approach to take with real data.

Our toy model included only 20 nuisance sources in total, whereas the LISA data
will contain $\mathcal{O}(10^4)$ individually resolvable Galactic binaries. We
expect that a small number of sources will still dominate the masking of any
given target, in which case the greedy ranking of Sec.~\ref{sec:selection} will
identify them for a refit. It could also be that no source dominates, but the
cumulative effect of many still leads to an appreciable masking, as the 16
crossing sources of our toy model illustrate in a mild form. This
would not go unnoticed, since the total masking estimated with
Eq.~\eqref{eq:sperpestimator} includes all sources, and the greedy algorithm
would still terminate, but the selected set could become large, making the
down-scaled reanalysis expensive. This is another point that should be tested
using the DDPC common datasets.

Finally, we also assumed stationary Gaussian noise with a known spectral density,
whereas in practice the noise will have to be inferred jointly with the
sources. Formally, the noise parameters then act as one more nuisance block,
but we expect them to matter less for the masking, because they couple in a different way to the
residual. A biased nuisance source is subtracted, so its imprint
$\Delta h$ is a waveform-like piece of the residual that the second step
projects onto the beyond-vacuum direction, linearly in $\thb$. Instead, a biased noise estimate cannot be
subtracted as it enters only through the weighting $C_n^{-1}$ of the inner
product, and at linear-signal order the expected Fisher cross-terms between
waveform and noise-model parameters vanish at the true parameters, because
they are proportional to the mean of the residual. What the noise sector does
see is the power of the residual, including that of the imprint left by
the biased subtraction, and this enters only at second order in the residual. It increases the
inferred spectral density in the frequency band where the residual lives, which does not
displace $\thb$ along a degeneracy but down-weights that band, lowering the
significance of every source there, including that of the effect itself. If
the inflation were frequency independent, all Fisher matrices would scale by
the same factor and the masking ratio would be unchanged. For a residual localised in frequency, the ratios have to be recomputed with the inflated spectral density,
although the fractional inflation is set by the residual's signal-to-noise
ratio per frequency bin, and is therefore small for effects that are not too strong. The population of unresolved sources below a certain threshold
can also be included in the analysis in the same way, either as an effective bias or as extra
confusion noise~\cite{Antonelli:2021vwg}.
Again, this is one more aspect we can study in more detail  in realistic mock datasets to ensure that by the time LISA flies, we have a robust inference strategy to do precision tests of General Relativity.

\acknowledgments

We thank the organisers and participants of the meetings on \textit{EMRI Search and Inference within the LISA Global Fit} - Part I (Paris) \& Part II (Singapore) for very fruitful discussions that helped construct and improve the ideas present in this manuscript. These meetings were supported by Université Paris Cité - National University of Singapore (Paris-NUS) research grant 2024-01-RESEARCH/UPCité-NUS (IdEx code ANR-18-IDEX-0001). We also thank Arnab Dhani, Rohit Chandramouli and Nicol\'as Yunes for insightful comments that helped improve
this manuscript. The work of F.D. is supported by the European Research Council (ERC) Advanced Grant \textit{De Tenebris} (grant agreement No. 101198798), funded by the European Union. Views and opinions expressed are however those of the author(s) only and do not necessarily reflect those of the European Union or the European Research Council Executive Agency. Neither the European Union nor the granting authority can be held responsible for them. J.~G. acknowledges support from the Simons Foundation International via Grant No. SFI-MPS- BH-00012593-06.

\appendix

\section{The masking inequality}
\label{app:lemma}

We prove here that the two terms of
Eq.~\eqref{eq:DeltaS} are separately non-negative, so that
$\mathcal{R}\in[0,1]$ and $\mathcal{R}_0\in[0,1]$, and that the weights $w_i$ of
Eq.~\eqref{eq:chi2pp} are real and lie in $[0,1]$.

Let us  write $A=\Fc{b}{cv}$ and $B=\Fc{b}{v}$, both symmetric, and assume
$A\succ0$. We only need to use that a symmetric
positive-definite matrix has a symmetric, invertible square root $B^{1/2}$, and that for a positive semi-definite matrix $M$, the matrix
$C^{T}MC$ is also positive semi-definite, since
$x^{T}C^{T}MCx=(Cx)^{T}M(Cx)$. First,
by Eq.~\eqref{eq:cGamma},
$B-A=\Gamma_{\mathrm{cb}\mid\mathrm{v}}^{T}\Fc{c}{v}^{-1}
\Gamma_{\mathrm{cb}\mid\mathrm{v}}$, so for any vector $x$
\begin{equation}
x^{T}(B-A)\,x=y^{T}\Fc{c}{v}^{-1}y\;\ge\;0 \, ,
\qquad
y=\Gamma_{\mathrm{cb}\mid\mathrm{v}}\,x \, ,
\label{eq:A1}
\end{equation}
because $\Fc{c}{v}$ is itself a conditional Fisher matrix and hence positive
definite. Then, consider the symmetric, positive-definite matrix
\begin{equation}
K= B^{-1/2}\,A\,B^{-1/2} \, .
\label{eq:Kdef}
\end{equation}
If $Ke=w\,e$ with $e\neq\bm 0$, then taking $x=B^{-1/2}e$,
\begin{equation}
w=\frac{e^{T}Ke}{e^{T}e}=\frac{x^{T}A\,x}{x^{T}B\,x}\;\in\;(0,1] \, ,
\end{equation}
by Eq.~\eqref{eq:A1} and $A\succ0$. Also, $K$ is similar to
$B^{-1}A=B^{-1/2}KB^{1/2}$ and to
$A^{1/2}B^{-1}A^{1/2}=A^{1/2}(B^{-1}A)A^{-1/2}$, and similar matrices have the same
eigenvalues, so the weights $w_i$ of Eq.~\eqref{eq:chi2pp} are real and
between $[0,1]$. The trace is also invariant for similar matrices
\begin{equation}
\Tr\bigl(B^{-1}A\bigr)=\Tr K=\sum_{i=1}^{N_{\mathrm b}}w_i\;\le\;N_{\mathrm b} \, ,
\end{equation}
which is the second term of Eq.~\eqref{eq:DeltaS} and gives
$\mathcal{R}_0\in[0,1]$. Then
\begin{equation}
A-AB^{-1}A=B^{1/2}\,K(\II-K)\,B^{1/2} \, .
\label{eq:KIK}
\end{equation}
The matrix $K(\II-K)$ is symmetric with eigenvalues $w_i(1-w_i)\ge0$, so
positive semi-definite, and so is the right-hand side of the equation above. Evaluated
at $\thb_{\mathrm{eff}}$, this is the first term of Eq.~\eqref{eq:DeltaS}.
Explicitly, writing $y=B^{1/2}\thb_{\mathrm{eff}}$ and $e_i$ for the orthonormal
eigenvectors of $K$,
\begin{equation}
\lambda^{\mathrm{GF}}-\lambda^{\mathrm{PP}}
=y^{T}K(\II-K)\,y
=\sum_i w_i(1-w_i)\,(e_i^{T}y)^{2}\;\ge\;0 \, ,
\label{eq:modewise}
\end{equation}
so $\mathcal{R}\in[0,1]$. Since
$\lambda^{\mathrm{PP}}/\lambda^{\mathrm{GF}}
=\sum_i w_i^{2}(e_i^{T}y)^{2}/\sum_i w_i(e_i^{T}y)^{2}$
is a weighted average of the $w_i$, we also get
$w_{\mathrm{min}}\le\mathcal{R}^{2}\le w_{\mathrm{max}}$, as stated in
Eq.~\eqref{eq:Rbracket}.

\pagebreak

\section{Vacuum injection: 2D posteriors}
\label{app:JointPosteriorNull}

We show here the marginalised 2D-posterior over the full parameter space of the target source of
interest for the vacuum injection, $\varepsilon_0=0$.

\begin{figure}[h]
\centering
\includegraphics[width=\textwidth]{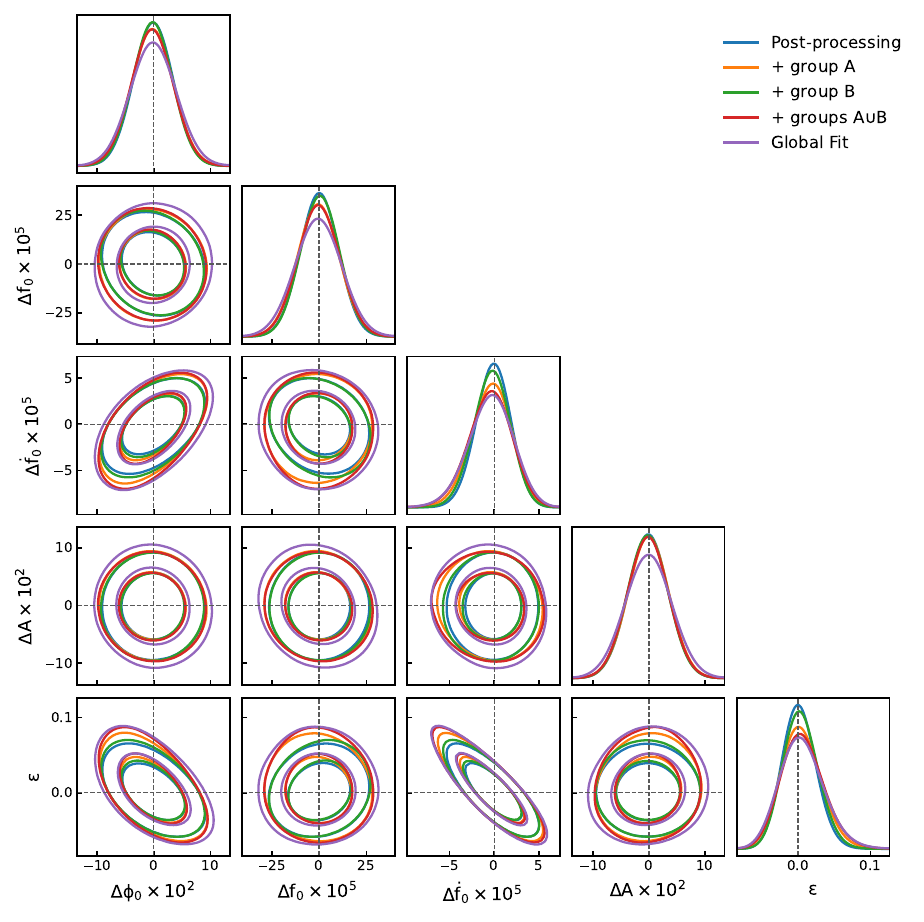}
\caption{Vacuum injection: marginalised posterior of the target source's
parameters for the five analyses of Sec.~\ref{sec:ToyNull}. Contours enclose
$68\%$ and $95\%$ of the posterior probability, and the injected values are
marked by the dashed lines. }
\label{fig:cornernull}
\end{figure}

\pagebreak

\section{Beyond-vacuum GR injection: 2D posteriors}
\label{app:JointPosteriorBGR}

The same posterior as in Appendix~\ref{app:JointPosteriorNull} for the beyond-vacuum GR injection, $\varepsilon_0=0.1$, and also for the groups A and B of the nuisance sources, for which we also show the first step of post-processing.

\begin{figure}[h]
\centering
\includegraphics[width=\textwidth]{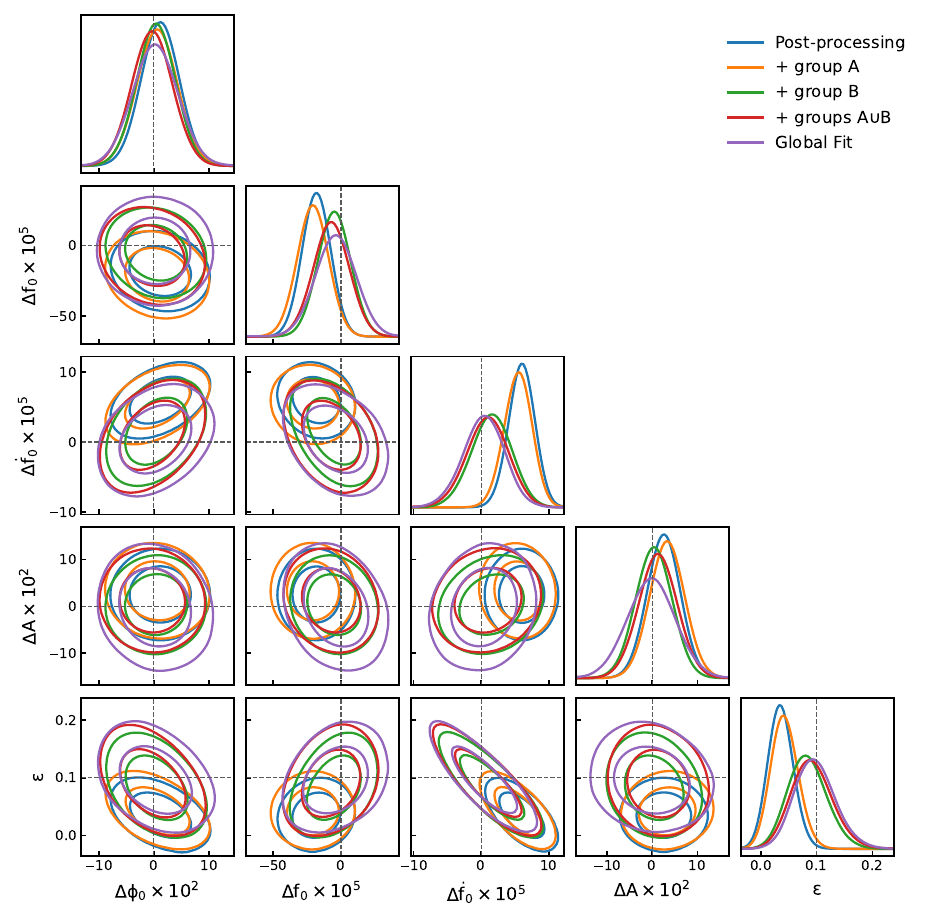}
\caption{Same as Fig.~\ref{fig:cornernull}, for the beyond-vacuum GR injection
$\varepsilon_0=0.1$. The post-processing recovery is biased from the truth
along the $\varepsilon$--$\dot f_0$--$f_0$ degeneracy, so the target's vacuum
parameters are biased together with $\varepsilon$. Freeing group B in the second step of post-processing (or both
groups) recovers almost the global-fit posterior, while freeing group A alone
leaves it essentially unchanged.}
\label{fig:cornerbv}
\end{figure}

\begin{figure}[h]
\centering
\includegraphics[width=\textwidth]{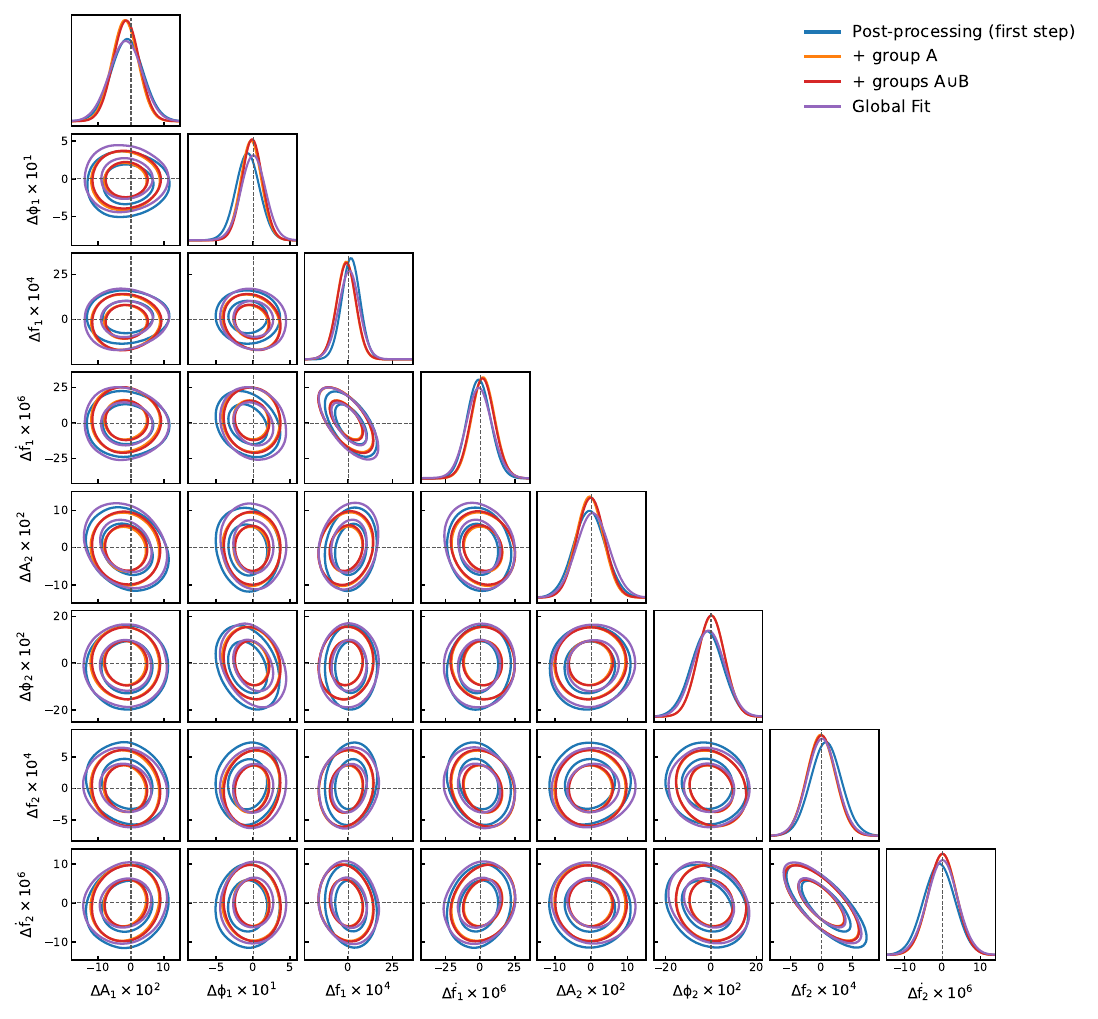}
\caption{Same as Fig.~\ref{fig:cornerbv} for the nuisance sources in group A, plotted as displacements from the injected values. Blue is the first step
of post-processing, the vacuum global fit, at whose best fit these sources are
frozen and subtracted in target-only post-processing. Group A absorbs little of
the effect at $\varepsilon_0=0.1$, so this posterior is displaced from the
injection by at most $0.5\sigma$. Once these
sources are refitted alongside the target they are recovered on their injected
values, to within $0.5\sigma$, with uncertainties around $10$--$15\%$ tighter
than the global-fit ones because the remaining nuisance sources stay fixed in the refit.}
\label{fig:cornerAbv}
\end{figure}

\begin{figure}[h]
\centering
\includegraphics[width=\textwidth]{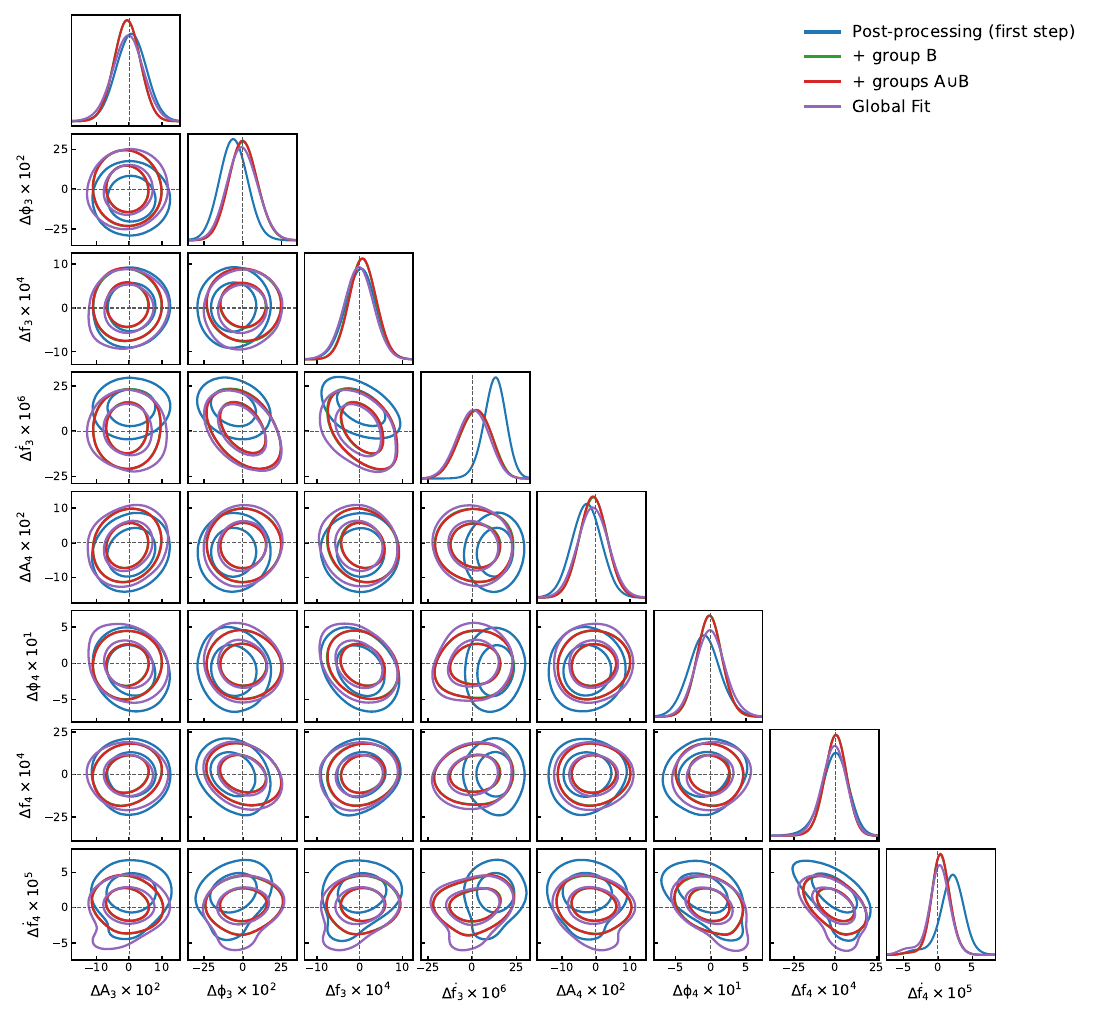}
\caption{Same as Fig.~\ref{fig:cornerAbv} for the nuisance sources in group B. Here the first step (blue) is visibly biased: the best-fit chirp rates of
sources $3$ and $4$ lie $2.4\sigma$ and $1.1\sigma$ from the injection, in units of
the first-step posterior width $\sigma$, and
their other parameters by up to $0.7\sigma$. These are the values subtracted in
target-only post-processing, while refitting group B with the target returns
them to within $0.3\sigma$ of the injection. The $+$ group B and $+$ groups
A$\cup$B posteriors coincide, so the former is hidden.}
\label{fig:cornerBbv}
\end{figure}

\bibliographystyle{JHEP}
\bibliography{biblio}

\end{document}